\pdfoutput=1
\documentclass[11pt,twoside,a4paper,cmspaper,final,collab]{cms-tdr}
\def\svnVersion{e9bc1c4}\def\svnDate{2024/07/19}\def\cmsCernNoTag{CERN-EP-2024-022}\def\cmsCernDate{\today}\def\cmsMessage{Published in Physical Review D as \href{http://dx.doi.org/10.1103/PhysRevD.110.012004}{\doi{10.1103/PhysRevD.110.012004}.}}
\begin{document}\cmsNoteHeader{EXO-22-017}

\newlength\cmsTabSkip\setlength{\cmsTabSkip}{1ex}
\newlength\cmsFigWidth
\ifthenelse{\boolean{cms@external}}{\setlength\cmsFigWidth{0.48\textwidth}}{\setlength\cmsFigWidth{0.9\textwidth}}
\ifthenelse{\boolean{cms@external}}{\providecommand{\cmsLeft}{upper\xspace}}{\providecommand{\cmsLeft}{left\xspace}}
\ifthenelse{\boolean{cms@external}}{\providecommand{\cmsRight}{lower\xspace}}{\providecommand{\cmsRight}{right\xspace}}
\newcommand{\hnl}{{\HepParticle{N}{}{}}\xspace}
\newcommand{\mhnl}{\ensuremath{m_\hnl}\xspace}
\newcommand{\VNl}{\ensuremath{V_{\hnl\Pell}}\xspace}
\newcommand{\VNlsq}{\ensuremath{\abs{\VNl}^2}\xspace}
\newcommand{\VNe}{\ensuremath{V_{\hnl \Pe}}\xspace}
\newcommand{\VNesq}{\ensuremath{\abs{\VNe}^2}\xspace}
\newcommand{\VNmu}{\ensuremath{V_{\hnl\PGm}}\xspace}
\newcommand{\VNmusq}{\ensuremath{\abs{\VNmu}^2}\xspace}
\newcommand{\VNtau}{\ensuremath{V_{\hnl\PGt}}\xspace}
\newcommand{\VNtausq}{\ensuremath{\abs{\VNtau}^2}\xspace}
\newcommand{\CP}{\ensuremath{CP}\xspace}
\newcommand{\ctau}{\ensuremath{c\tau_{0}}\xspace}
\newcommand{\keVns}{\ensuremath{\text{ke\hspace{-.08em}V}}\xspace}
\newcommand{\dphilep}{\ensuremath{\Delta\phi_\text{lep}}\xspace}
\newcommand{\nhits}{\ensuremath{N_{\text{hits}}}\xspace}
\newcommand{\nstations}{\ensuremath{N_{\mathrm{stations}}}\xspace}
\newcommand{\zmm}{\ensuremath{\PZ\to\PGm\PGm}\xspace}

\cmsNoteHeader{EXO-22-017}
\title{Search for long-lived heavy neutral leptons decaying in the CMS muon detectors in proton-proton collisions at \texorpdfstring{$\sqrt{s}=13\TeV$}{sqrt(s) = 13 TeV}}

\date{\today}

\abstract{
A search for heavy neutral leptons (HNLs) decaying in the CMS muon system is presented. A data sample is used corresponding to an integrated luminosity of 138\fbinv of proton-proton collisions at $\sqrt{s}=13\TeV$, recorded at the CERN LHC in 2016--2018. Decay products of long-lived HNLs could interact with the shielding materials in the CMS muon system and create hadronic and electromagnetic showers detected in the muon chambers. This distinctive signature provides a unique handle to search for HNLs with masses below 4\GeV and proper decay lengths of the order of meters. The signature is sensitive to HNL couplings to all three generations of leptons. Candidate events are required to contain a prompt electron or muon originating from a vertex on the beam axis and a displaced shower in the muon chambers. No significant deviations from the standard model background expectation are observed. In the electron (muon) channel, the most stringent limits to date are set for HNLs in the mass range of 2.1--3.0 (1.9--3.3)\GeV, reaching mixing matrix element squared values as low as $8.6\,(4.6) \times10^{-6}$.
}

\hypersetup{
  pdfauthor={CMS Collaboration},
  pdftitle={Search for long-lived heavy neutral leptons decaying in the CMS muon detectors in proton-proton collisions at sqrt(s) = 13 TeV},
  pdfsubject={CMS},
  pdfkeywords={CMS, LLP, long-lived particle, HNL, muon system}
}

\maketitle

\section{Introduction}
\label{sec:intro}

The observation of neutrino oscillations~\cite{Super-Kamiokande:1998kpq,SNO:2002tuh, KamLAND:2002uet} provides experimental evidence for nonzero neutrino masses~\cite{Bilenky:2016pep}.
Cosmological considerations~\cite{RoyChoudhury:2019hls,Ivanov:2019hqk} and direct measurements~\cite{Formaggio:2021nfz} imply that the neutrino masses are much smaller than both the masses of other standard model (SM) fermions and the vacuum expectation value of the Higgs field, hinting at a different mechanism for generating masses of neutrinos compared with that for the masses of charged fermions.

One possible explanation of the nonzero neutrino masses is the existence of heavy neutral leptons (HNLs) with right-handed chirality, giving rise to the gauge-invariant mass terms for the SM neutrinos through the ``seesaw'' mechanism~\cite{Minkowski:1977sc, Yanagida:1979as, Gell-Mann:1979vob, Glashow:1979nm, Mohapatra:1979ia, Schechter:1980gr, Shrock:1980ct, Cai:2017mow}.
The HNLs are singlets with respect to the SM gauge groups and therefore do not interact with SM particles through the electroweak or strong interactions, but can be produced through mixing with the SM electron, muon, and $\PGt$ neutrinos~\cite{Maki:1962mu, Pontecorvo:1967fh, Das_2016, Das_2017, cottin2023long}.
There can be either distinct HNL particles and antiparticles [(pseudo) Dirac type], or the HNL can be its own antiparticle [Majorana type].
An HNL is characterized by its mass, \mhnl, and its mixing matrix elements \VNl, with $\ell=\Pe,\PGm,\PGt$, respectively.
The HNL lifetime is inversely proportional to $\mhnl^5 \VNlsq$~\cite{Bondarenko:2018ptm}, and HNLs can be macroscopically long-lived for sufficiently low values of mixing matrix elements.
Models with HNLs are well motivated because they can explain the baryon asymmetry of the universe through \CP violation in the HNL system~\cite{Fukugita:1986hr, Chun:2017spz}, provide a dark matter candidate~\cite{boyarsky2019sterile}, and explain the observed anomalous magnetic moment of the muon~\cite{Cirigliano_2021,Muong-2:2023}.

Past searches for HNLs have covered a wide range of masses ranging from a few \keVns to several \TeVns~\cite{BEBC,CHARM:1985nku,delphi1997search,NuTeV,Belle,Deppisch:2015qwa,Beacham:2019nyx}.
At the CERN LHC, searches conducted by the ATLAS, CMS, and LHCb Collaborations~\cite{ATLAS:2019kpx,ATLAS:2022atq,CMS:2018iaf,CMS:2018jxx,CMS:2022fut,LHCb:2020wxx,LHCb:2020akw} have targeted both prompt and long-lived HNL decays.
The search presented in this paper focuses on HNLs with masses between 1~and~4\GeV.
In this region, the most stringent prior limits for the muon (electron) mixing parameter \VNmusq (\VNesq) have been set by the BEBC, CHARM and NuTeV experiments~\cite{BEBC,CHARM:1985nku,NuTeV} below about 1.9\,(2.1)\GeV, and by the CMS searches~\cite{CMS:2022fut,CMS:2023jqi} for masses above 1.9\,(2.1)\GeV.
The most stringent prior limits for the $\PGt$ mixing parameter \VNtausq have been set by the DELPHI and CHARM experiments~\cite{delphi1997search,CHARM:1985nku}.
This search aims to extend the discovery reach towards the lowest values of $\VNlsq$ achieved in the 1--4\GeV mass range.

\begin{figure}[!htb]
  \centering
  \includegraphics[width=0.48\textwidth]{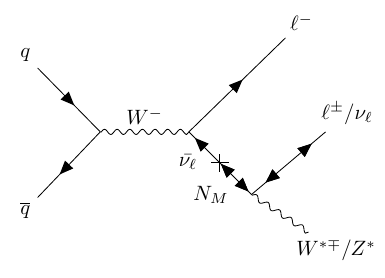}
  \includegraphics[width=0.48\textwidth]{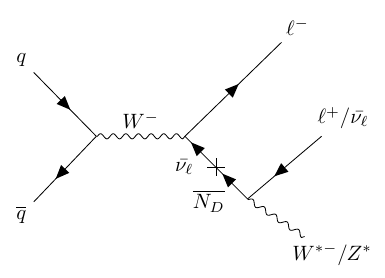}
  \caption{
    Feynman diagrams for the production of a Majorana HNL $N_M$ (\cmsLeft) and a Dirac HNL $\overline{N_D}$ (\cmsRight) via a \PWm boson decay and through its mixing with an SM neutrino of the same flavor.
    The prompt lepton from the \PWm boson serves as a clean signature for triggering, whereas the decay products of the HNL are reconstructed as a cluster of muon detector hits.
  }
  \label{fig:Feynman}
\end{figure}

In this paper, a search for Dirac and Majorana HNLs with mean proper decay lengths (\ctau) in the range of 0.1--10\unit{m} is performed.
A data sample collected by the CMS experiment in 2016--2018 and corresponding to an integrated luminosity of 138\fbinv is used.
An HNL that decays via a virtual \PW boson can result in a final state with two charged leptons and a neutrino, or one lepton and two quarks.
Figure~\ref{fig:Feynman} shows the Feynman diagram for the production of an HNL via \PW boson decay, where the prompt lepton from the \PW boson decay originates from a vertex on the beam axis and serves as a clean signature for triggering.
We search for HNL decays occurring within the muon detector system.
The hadrons and electrons from the HNL decays, including those resulting from an intermediate $\PGt$ lepton, could produce a particle shower as they interact with the steel shielding material of the muon detectors.
This process results in the striking signature of a localized high-multiplicity cluster of muon detector hits, referred to as a muon detector shower (MDS) object.
Because of the macroscopic distance from the primary interaction point to the muon detectors and its large volume of geometric acceptance, this signature is sensitive to a large class of models involving long-lived particles (LLPs).
It is uniquely sensitive to HNLs with \ctau in the range 0.1--10\unit{m}, extending the search sensitivity to unprecedentedly low mixing parameter values in the \mhnl range of 1--3\GeV.
We follow an analysis strategy for MDS objects that is similar to one used by CMS in Ref.~\cite{CMS:2021juv}.

The outline of the paper is as follows: a brief description of the CMS detector is given in Section~\ref{sec:detector}.
The data set and the simulated samples of events are detailed in Section~\ref{sec:samples}, followed by a description of the event selection and categorization in Section~\ref{sec:selection}.
A description of the backgrounds and the method used to estimate them is given in Section~\ref{sec:bkg}.
Sources of systematic uncertainty affecting the expected numbers of signal and background events are discussed in Section~\ref{sec:systematics}.
The results and their interpretations are presented in Section~\ref{sec:results}, followed by a summary of the paper in Section~\ref{sec:summary}.
The numerical results of this paper can be found in its HEPData record~\cite{hepdata}.

\section{The CMS detector}
\label{sec:detector}
The central feature of the CMS detector is a superconducting solenoid of 6\unit{m} internal diameter, providing a magnetic field of 3.8\unit{T}.
Within the solenoid volume are a silicon pixel and strip tracker, a lead tungstate crystal electromagnetic calorimeter (ECAL), and a brass and scintillator hadron calorimeter (HCAL), each composed of a barrel and two endcap sections.
The ECAL consists of 75,848 lead tungstate crystals, which cover $\abs{\eta} < 1.48 $ in the barrel region and $1.48 < \abs{\eta} < 3.00$ in the two endcap regions.
The HCAL is composed of cells of width 0.087 in $\eta$ and azimuth ($\phi$, in radians) for the region $\abs{\eta} < 1.74$, progressively increasing to a maximum of 0.174 for larger values of $\abs{\eta}$, along with the forward calorimeters extending the $\eta$ coverage provided by the barrel and endcap detectors.

Muons are measured in the pseudorapidity range of $\abs{\eta} < 2.4$, with detection planes embedded in the steel flux-return yoke outside the solenoid and made using three technologies: drift tubes (DTs), cathode strip chambers (CSCs), and resistive-plate chambers (RPCs).

The barrel DT detectors cover $\abs{\eta} < 1.2$ and are organized into four layers (``stations"), labeled MB1 to MB4.
The stations are located approximately 4, 5, 6, and 7\unit{m} in radial distance ($r$) from the interaction point.
Each station is a concentric ring of chambers, interleaved with the layers of the steel flux-return yoke of the solenoid and is divided along the beamline axis ($z$) into five wheels.
In the first three stations, every DT chamber consists of three ``superlayers'' (SLs), each comprising four staggered layers of parallel DT cells, for a total of 12 layers.
The innermost and outermost SLs measure the hit coordinate in the $r\phi$ plane, and the central SL measures in the $z$ direction, along the beamline.
The fourth station (MB4) contains only two SLs measuring the hit position in the $r\phi$ plane.
The DT cells are designed to provide a uniform electric field, such that the position of a traversing charged particle can be inferred from the measured arrival time and the constant drift velocity.
Individual hits have a resolution of about 600\mum which can be improved to about 260\mum by incorporating information from hits in the same SL~\cite{CMS:2009lnj}.

The CSC detectors cover $0.9 < \abs{\eta} <  2.4$ and comprise four stations in each endcap, labeled ME1 to ME4.
Each station is a ring of chambers interleaved between two layers of steel-flux return yoke at approximately the same value of $z$.
The stations are located approximately 7, 8, 9.5, and 10.5\unit{m} from the interaction point along the beamline axis on both ends of the detector.
Each chamber consists of six layers containing cathode strips along the radial direction and anode wires perpendicular to the central strip.
Position and timing measurements of traversing charged particles are extracted from the electrical signals on the anode wires and the cathode strips in each chamber, for which the typical resolutions are 400--500\mum and 5\unit{ns}.

The RPC detectors are placed alongside the DT and CSC detectors and are primarily designed to provide timing information for the muon trigger.

Events of interest are selected using a two-tiered trigger system.
The first level, composed of custom hardware processors, uses information from the calorimeters and muon detectors to select events at a rate of around 100\unit{kHz} within a fixed latency of about 4\mus~\cite{CMS:2020cmk}.
The second level, known as the high-level trigger, consists of a farm of processors running a version of the full event reconstruction software optimized for fast processing, and reduces the event rate to around 1\unit{kHz} before data storage~\cite{CMS:2016ngn}.
A more detailed description of the CMS detector, together with a definition of the coordinate system used and the relevant kinematic variables, is reported in Ref.~\cite{CMS:2008xjf}.

\section{Simulated samples}
\label{sec:samples}
{\tolerance=800
  Simulated samples of Drell-Yan produced $\PZ$+jets are used to validate the MDS object reconstruction efficiency.
  They are produced using \POWHEG~2.0~\cite{POWHEG1,POWHEG2,POWHEG3} and normalized to next-to-next-to-leading-order (NNLO) precision~\cite{Gavin:2010az, Gavin:2012sy}.
  Signal samples for HNL production are generated at leading order (LO) with the Monte Carlo (MC) generator \MGvATNLO 2.3.3~\cite{Alwall_2014,Alva_2015,Degrande_2016,Pascoli_2019}.
  This matrix element level calculation includes only the \PW-channel production of HNLs with up to two additional partons.
  The simulated events are interfaced with \PYTHIA8.226~\cite{Sj_strand_2015} to simulate the parton shower and hadronization of partons, and the underlying event description.
  \par}

The branching fraction of \PW boson decays into an HNL plus a lepton is proportional to the mixing parameter \VNlsq.
The signatures considered in this search are sensitive to mixing with \PGne, \PGnGm, or \PGnGt since the decay products of the HNL in the muon detector system are reconstructed as particle showers.
The final states consist of one prompt lepton (\Pe or \PGm) from the \PW boson decay and HNL decay products.
To account for higher order effects, the cross sections of the signal samples are scaled to the NNLO \PW boson production cross section, which is $20\,510\pm770$\unit{pb}~\cite{FEWZ} including the branching fraction to one lepton flavor, multiplied by the square of the HNL mixing matrix elements.
The branching ratios to all three lepton generations are assumed to be the same.
To obtain a more accurate description of the boost of the HNL signal, \PW-boson-\pt-dependent corrections based on an MC sample generated with \textsc{DYTurbo}1.3.2 at next-to-next-to-next-to-leading-logarithm and next-to-NNLO accuracy~\cite{DYTURBO,Camarda_2021} are computed and applied as an event-by-event weight.

Signals corresponding to both Dirac and Majorana HNLs are considered in this search, for \ctau values in the range of 0.1--10\unit{m}.
Lepton number violating (LNV) decays are only possible for Majorana HNLs, whereas lepton number conserving (LNC) decay channels are available for both Majorana and Dirac HNLs.
This distinction implies that the widths of the Dirac HNL are exactly half of that of the Majorana HNLs for the same \mhnl and mixing matrix; therefore, the lifetime of the Dirac HNLs will be twice the Majorana ones~\cite{Pascoli_2019}.
In this search, the charged lepton from the HNL decay is either not detected or its charge is not measured, resulting in identical detector response between LNV and LNC events.
Therefore, we can predict the signal yields of a simulated Majorana sample as those from a Dirac HNL sample with twice the lifetime.

Generated events are processed through a simulation of the detector geometry and response using \GEANTfour~\cite{AGOSTINELLI2003250}.
The same reconstruction software is applied to both data and simulated events.
Simulated events include an expected distribution of the number of additional $\Pp\Pp$ interactions within the same or nearby bunch crossings (pileup).
All generated events are weighted such that the distribution of the number of collisions per bunch crossing matches the one observed in data, with an average of approximately 23 (32) interactions per bunch crossing~\cite{CMS-LUM-17-003, CMS-PAS-LUM-17-004, CMS-PAS-LUM-18-002} in 2016 (2017--2018).
The underlying Pythia event tune CUETP8M1~\cite{Khachatryan:2015pea} (CP5~\cite{CP5}) and NNPDF3.0~\cite{Ball:2014uwa} (3.1~\cite{Ball:2017nwa}) parton distribution functions are used to simulate all samples for the 2016 (2017--2018) data taking period.

The HNL signal samples are generated for \mhnl in the range from 1 to 4\GeV and \ctau in the range from 0.1 to 10\unit{m}.
Since only a discrete set of lifetimes was generated, the signal predictions for intermediate lifetimes were estimated by performing the reweighting procedure described below.
The simulated HNLs are assumed to couple exclusively to one of the three SM neutrino families.
The signal production cross sections and lifetimes are determined by the values of \mhnl and \VNlsq, which in turn determine the signal acceptance and reconstruction efficiency.
Thus, for a fixed value of \mhnl, a simple cross section rescaling is not sufficient to correctly reproduce the behavior of other HNLs with the same \mhnl but different \VNlsq.
To emulate an HNL signal sample with a specific value of \VNlsq (and thus \ctau), we thus apply per-event weights to the events, such that the HNL lifetime distribution (taken before the parton shower and detector simulation) matches the predicted distribution for the chosen \VNlsq value.
For signal scenarios with \mhnl between the masses of the simulated samples, we estimate the signal yield by interpolating between the predictions from the simulated samples at nearby \mhnl using the fact that the acceptance of the muon system differs only through the Lorentz boost factor.

\section{Event reconstruction and selection}
\label{sec:selection}

The objects in each event are reconstructed using the particle-flow (PF) algorithm~\cite{Sirunyan:2017ulk}, which aims to identify each individual particle in an event as an electron, photon, muon, charged or neutral hadron, with an optimized combination of information from the various elements of the CMS detector.
The resulting particles are referred to as PF candidates.
Each electron is identified as a charged-particle track that extrapolates to an ECAL energy cluster and any nearby ECAL clusters that are consistent with possible bremsstrahlung photons~\cite{CMS:2020uim}.
The electron momentum is estimated by combining the energy measurement in the ECAL with the momentum measurement in the tracker.
Muons are identified as tracks in the central tracker consistent with either tracks or several hits in the muon system, and associated with calorimeter deposits compatible with the muon hypothesis~\cite{CMS:2018rym}.
Photons are identified as ECAL energy clusters not linked to the extrapolation of any charged-particle trajectory to the ECAL~\cite{CMS:2015myp}.
The energies of photons are obtained from the ECAL measurement.
Charged hadrons are identified as charged-particle tracks neither identified as electrons, nor as muons.
The energy of each charged hadron is determined from a combination of the track momentum and the corresponding ECAL and HCAL energies, corrected for the response function of the calorimeters to hadronic showers.
Finally, a neutral hadron is identified as HCAL energy clusters not linked to any charged-hadron trajectory, or as a combined ECAL and HCAL energy excess with respect to the expected charged-hadron energy deposit.
The energy of each neutral hadron is obtained from the corresponding corrected ECAL and HCAL energies.

To identify the electrons and muons resulting from the decays of the \PW bosons, we impose additional selection requirements to enhance the signal purities of these objects.
For each electron candidate, requirements are imposed on the shape of the electromagnetic shower in the ECAL, the quality of the matching between the track trajectory and the ECAL shower, and isolation from additional particles near the candidate.
A tight working point~\cite{CMS:2020uim} for this electron identification is used, which has an average efficiency of 70\% and a misidentification rate of about 2\%.
We consider electron candidates with transverse momenta $\pt > 30\GeV$ and $\abs{\eta} < 2.5$; a more stringent requirement of $\pt > 35\GeV$ is imposed for the data collected in 2017--2018 to match the increase in the \pt threshold of the single electron trigger from 27 to 32\GeV.
For muon candidates, requirements are imposed on the quality of the track in the silicon tracker, the global compatibility of the hits in the silicon tracker and muon detector comprising the muon candidate, and isolation from additional particles near the candidate.
The tight working point~\cite{CMS:2018rym} is used, which has an efficiency above 95\% and a misidentification rate of about 0.1\% for hadrons.
We consider muon candidates with $\pt > 25\GeV$ and $\abs{\eta} < 2.4$; a more stringent requirement of $\pt > 28\GeV$ is imposed for the data collected in 2017 to match the increase in the \pt threshold of the single-muon trigger from 24 to 27\GeV.

In this search, jets are used to veto muon detector shower cluster objects to suppress the punch-through jet background.
Jets are reconstructed by clustering PF candidates using the anti-\kt algorithm~\cite{Cacciari:2008gp, Cacciari:2011ma} with a distance parameter $\Delta R = \sqrt{\smash[b]{(\Delta\eta)^2 + (\Delta\phi)^2}} = 0.4$.
Jets are required to have $\pt > 10\GeV$ and $\abs{\eta} < 3.0$.
Pileup can contribute additional tracks and calorimetric energy depositions to the jet momentum.
To mitigate this effect, charged particles identified to be originating from pileup vertices are discarded and an offset correction is applied to account for remaining contributions~\cite{CMS:2020ebo}.
Jet energy corrections are derived from simulation studies so that the average measured energy of jets becomes identical to that of particle-level jets~\cite{Khachatryan:2016kdb}.

The missing transverse momentum vector \ptvecmiss is computed as the negative vector \pt sum of all the PF candidates in an event, and its magnitude is denoted as \ptmiss~\cite{CMS:2019ctu}.
The jet energy corrections are propagated into the computation of \ptmiss.
The primary vertex (PV) is taken to be the vertex corresponding to the hardest scattering in the event, evaluated using tracking information alone, as described in Section 9.4.1 of Ref.~\cite{CMS-TDR-15-02}.

We select data events triggered by the single-electron or single-muon triggers and require the events to have exactly one prompt electron or prompt muon candidate, respectively, satisfying the identification and isolation criteria described above.
To further suppress the background from SM events composed uniquely of jets produced through the strong interaction, referred to as quantum chromodynamics (QCD) multijet events, we require $\ptmiss > 30\GeV$.
The above requirements are designed to select events with an HNL candidate produced in a \PW boson decay.
Because the constituents of MDS clusters are not included in the \ptmiss calculation, typical HNL signal events would satisfy this \ptmiss selection.

\subsection{Muon detector shower clusters}
\label{sec:clusterSelection}
For LLPs that decay within or just prior to the DT and CSC muon detectors, the material in the steel-flux return yoke structure can induce a particle shower, creating a geometrically localized and isolated cluster of detector hits.
In CSC chambers, the detector hits are reconstructed by combining the signal pulses from the anode wires and the cathode strips, forming a point on a two-dimensional plane in each chamber layer.
In DT chambers, the detector hits are reconstructed from the signal pulses from the anode wires at the center of the DT cells.
Since the DT hits only provide measurement in either the $\phi$ or $z$ dimension, the DT hit position is assumed to be at the center of each DT chamber in the orthogonal direction.
Together with the position of the chambers, each CSC and DT hit is assigned a three-dimensional coordinate in space.
We cluster the CSC and DT hits based on their $\eta$ and $\phi$ coordinates using the Density-Based Spatial Clustering of Applications with Noise (DBSCAN) algorithm~\cite{dbscan}, which is a commonly used density-based clustering algorithm that is robust against outliers.
A distance parameter of 0.2 and a minimum number of hits per cluster (\nhits) of 50 is used in the DBSCAN algorithm.
We choose a minimum of 50 hits to avoid misidentifying a minimum ionizing muon as a cluster object.
A muon is expected to produce a maximum of 24 and 44 hits in the CSC and DT detectors, respectively.
The centroid of the cluster is taken to be the geometric center of the hits in the cluster.
These clusters, which we refer to as MDS objects, provide a powerful signature to distinguish LLP signal events from background events.

The efficiency for MDS objects to be reconstructed depends on whether the LLP decays primarily to hadrons, including the hadronic decays of $\PGt$ leptons, or to electrons and photons.
Typically, more cluster hits are produced by hadrons, resulting in a higher MDS reconstruction efficiency for LLPs that decay to hadrons as compared with LLPs that decay to electrons and photons.
The reconstruction efficiency is about 80 (30--45)\% for hadronic (electron and photon) decays, depending on whether the MDS is reconstructed in the CSCs or DTs.
Muons that are produced from HNL decays do not produce a particle shower and are not reconstructed.
Further details of the efficiency dependence on LLP decay location can be found in reference~\cite{CMS:2021juv}.

Energetic particles produced in jet hadronization or from secondary material interactions can produce MDS clusters by traversing the shielding material between the tracking volume and the muon detectors without being stopped.
To suppress this process, known as the punch-through jet background, we veto any MDS clusters in the DT (CSC) detectors whose centroid lies within $\Delta R=0.4$ of a jet with $\pt > 20$ (10)\GeV and $\abs{\eta} < 3.0$.
To suppress clusters produced by muon bremsstrahlung, we impose a tighter veto on any MDS clusters in the DT (CSC) detectors whose centroid lies within $\Delta R=0.8$ of a muon candidate with $\pt > 10$ (20)\GeV.
The \pt thresholds for the jet and muon vetoes were chosen to optimize the signal-to-background discrimination and differ between the DT and CSC detectors because of the different amounts of shielding present.
Because MDS cluster hits do not enter the \ptvecmiss calculation, the MDS cluster tends to align with the \ptvecmiss and the momentum of the LLP in signal events.
We leverage this feature to validate the background estimation method discussed in Section~\ref{sec:bkg}.
Muons with trajectories pointing toward two regions of the detector near the cavern chimneys have highly reduced reconstruction efficiencies because these regions are occupied by cables and other support structures resulting in a smaller detector coverage.
As a result, the rate of background MDS clusters produced by muon bremsstrahlung and passing the muon veto is significantly increased.
To suppress these backgrounds, we veto any clusters whose centroid is within $\Delta R =0.3$ of the following two locations in $\eta$ and $\phi$ coordinates: ($\eta = 0.3$, $\phi = 1.70$) or ($\eta = -0.3$, $\phi = 1.15$), corresponding to the chimney locations.

The rate of MDS clusters caused by punch-through jets is significantly larger for clusters with hits in the muon detector station closest to the interaction point because of the reduced amount of shielding material.
Furthermore, any hits observed in the stations between the interaction point and the station containing the majority of the cluster hits are indicative of a punch-through jet because signal LLPs are neutral and therefore do not produce any hits before they decay.
To further suppress such punch-through jets, we veto any MDS clusters in the CSCs whose centroid lies within $\Delta R=0.4$ of any hits observed in the two innermost rings of the ME1 station (ME1/1 and ME1/2), or in the RPCs located immediately next to ME1/2, or any segments in the MB1 station.
Segments are required in the MB1 station instead of single hits because of the larger rate of noise in the DT detectors, which would produce a larger signal inefficiency for such a veto.
For MDS clusters in the DTs, we veto any cluster whose centroid lies within $\Delta R=0.5$ of two or more hits in the MB1 station.
We also require no more than 8 MB1 hits in the adjacent wheels within $\Delta\phi < \pi/4$.
As a result of the above vetoes, only clusters located beyond the station closest to the interaction point are accepted.

After the jet and muon vetoes, the dominant background consists of clusters from pileup, including interactions not in the same bunch crossing as the PV, known as out-of-time (OOT) pileup.
To eliminate the background clusters from OOT pileup, we reconstruct a characteristic time for the clusters in both the CSC and DT detectors.
For the CSC clusters, the cluster time is reconstructed from the mean time of the hits comprising the cluster, where the time of each hit is calibrated to result in a distribution centered at zero for the triggering bunch crossing.
Note that the adjacent bunch crossings occur at $\pm$25\unit{ns} with this definition.
We require that the CSC cluster time ($t_\mathrm{cluster}^\mathrm{CSC}$) is between -5.0 and 12.5\unit{ns} and that the root-mean-square spread of the timestamps of hits comprising the cluster ($t_\mathrm{spread}$) is less than 20\unit{ns}.
Because hits in the DT detectors do not have a time measurement that is independent of their position measurement, we require that DT clusters coincide with at least one hit in an RPC detector in the same wheel and within $\Delta\phi<0.5$ of the DT cluster centroid.
The time measurements of each matching RPC hit are matched to discrete bunch crossing times.
The most frequently occurring bunch crossing time among the matching RPC hits is defined as the DT cluster time ($t_\mathrm{cluster}^\mathrm{DT}$).
The DT cluster time is required to coincide with the bunch crossing corresponding to the PV.

Finally, additional identification requirements, developed in Ref.~\cite{CMS:2021juv}, are imposed on CSC clusters.
They are required to satisfy successively more central $\abs{\eta}$ requirements as the number of CSC stations containing hits (\nstations) and the distance between the station and the primary interaction point decrease.
The $\abs{\eta}$ requirements are
\begin{itemize}
  \item $\abs{\eta}<1.9$ if $\nstations>1$,
  \item $\abs{\eta}<1.8$ if $\nstations=1$ and the cluster is in station 4,
  \item $\abs{\eta}<1.6$ if $\nstations=1$ and the cluster is in station 2 or station 3, and
  \item $\abs{\eta}<1.1$ if $\nstations=1$ the cluster is in station 1.
\end{itemize}
This CSC cluster identification algorithm has $\sim$60\% efficiency and suppresses the background by a factor of 8.

For a certain period of data taking, anomalous detector noise in specific components of the DT detectors produced a high rate of background MDS clusters.
To suppress this noise-induced background, we veto MDS clusters comprising hits detected in those particular DT chambers taken during the affected time period.
The amount of data rejected by this veto corresponds to less than 0.1\% of the total integrated luminosity.

After applying all the cluster vetoes, we require the events to contain at least one CSC or DT cluster with $\nhits > 50$.
The minimum number of hits is chosen to exceed the number of hits that a muon is expected to create in either CSC or DT detectors.
The presence of an MDS cluster passing the associated vetoes and identification criteria suppresses SM background by a factor exceeding $10^7$, whereas typical signal efficiencies are 25--35\%.

\section{Background estimation}
\label{sec:bkg}
After the event selections described in Section~\ref{sec:selection}, two types of background events remain, in which the MDS cluster can be muon-induced or non-muon-induced.
The non-muon-induced background involves a hard scattering process that produces a prompt lepton, and low momentum hadrons from pileup or an underlying event generate an MDS cluster.
The dominant component of this background comes from \PW production, and subdominant contributions arise from  QCD, \ttbar, or diboson production.
The muon-induced background comes from \zmm events, in which one of the two muons from the \PZ boson undergoes bremsstrahlung in the muon detector and produces an MDS cluster back to back with the other prompt muon, mimicking the configuration of a signal event.
For the non-muon-induced background, which is present in both the prompt-muon and prompt-electron categories, we use an ``ABCD" (matrix) method, which requires two variables that discriminate between signal and background and are independent of one another for the background.
For the \zmm background, which is present only in the prompt-muon categories, the background estimation is derived from dedicated control regions in data.

\begin{figure*}[!htb]
  \centering
  \includegraphics[width=0.9\textwidth]{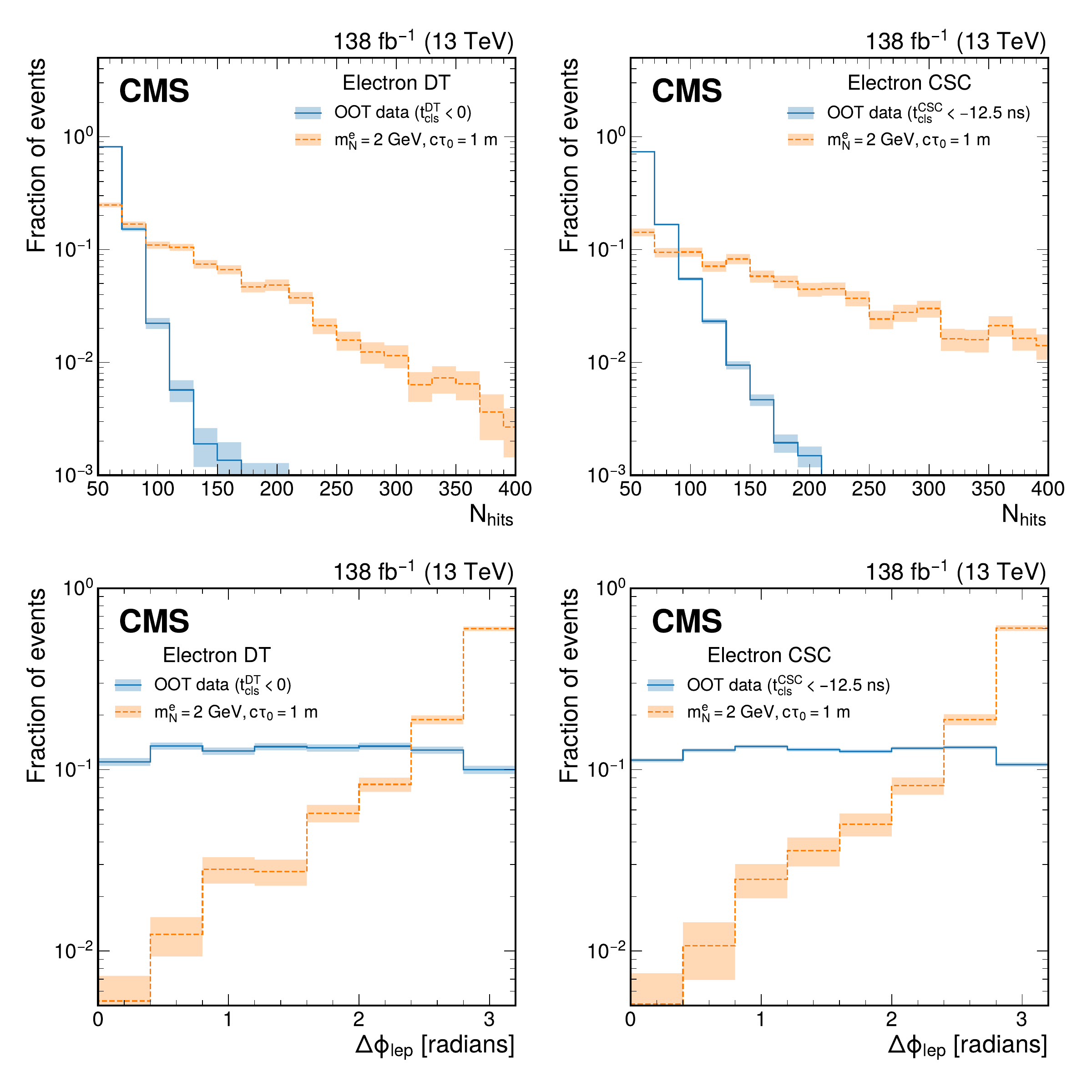}
  \caption{
    Distribution of \nhits (upper) and \dphilep (lower) for DT clusters (left) and CSC clusters (right).
    Signal distributions of a Majorana HNL with $\mhnl=2\GeV$ and $\ctau=1\unit{m}$ are compared with the OOT background distributions selected with $t_\mathrm{cluster}^\mathrm{DT}$ matched to bunch crossings earlier than the PV for DT clusters and $t_\mathrm{cluster}^\mathrm{CSC} <-12.5\unit{ns}$ for CSC clusters.
    The centroids of the clusters in the signal events are required to be within $\Delta R=0.4$ of the HNL's direction.
    The distributions are normalized to unit area.
    The shapes of the distributions shown are similar for the electron and muon channels.
  }
  \label{fig:distribution}
\end{figure*}

In the ABCD method, we select the two independent variables to be (i) the azimuthal angle between the prompt lepton and the cluster centroid (\dphilep) and (ii) \nhits.
For non-muon-induced events, \dphilep is uniformly distributed and is independent of \nhits, because the cluster and lepton are produced from two independent processes.
Figure~\ref{fig:distribution} shows the shapes of the \nhits and \dphilep distribution for signal and background.

\begin{figure}[!htb]
  \centering
  \includegraphics[width=0.48\textwidth]{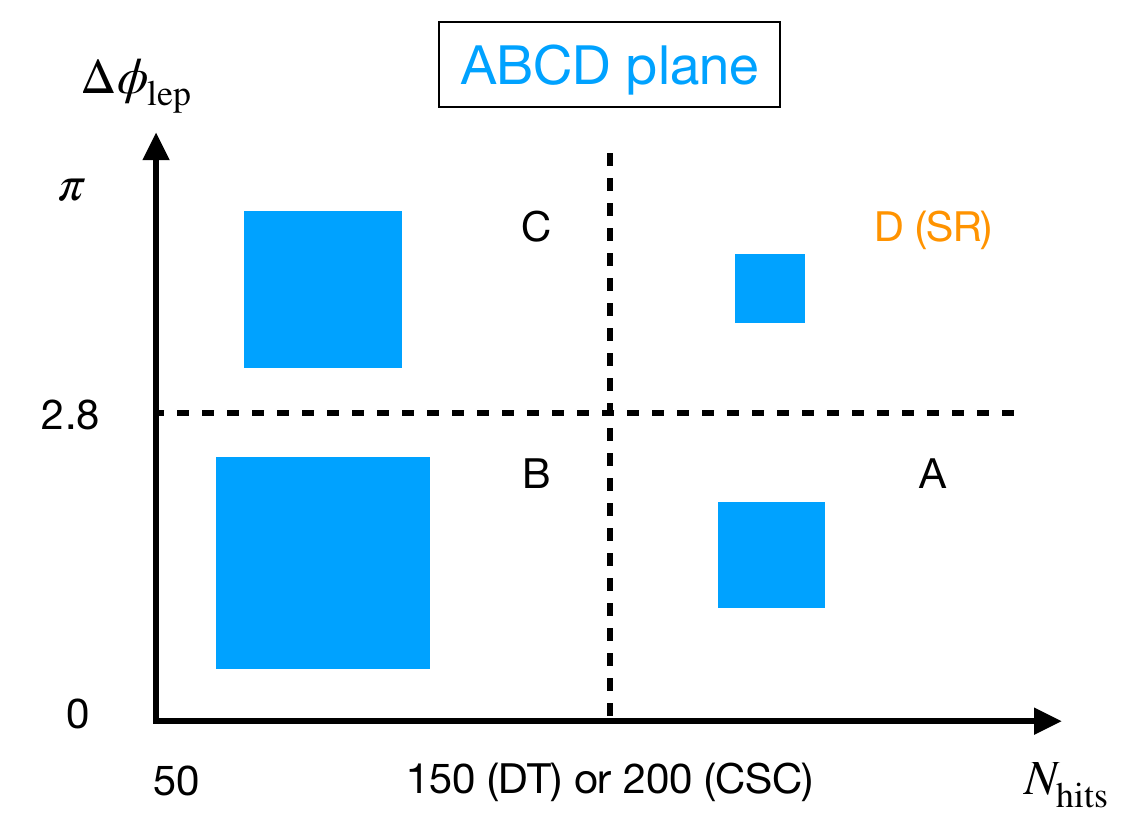}
  \caption{
    Definition of the ABCD plane.
    The area of the blue squares illustrates the relative amount of expected events in each of the bins, with bins B and C having the majority of the event yields.
    Bin D is the signal region.
  }
  \label{fig:ABCDplane}
\end{figure}

Two separate requirements, one on each variable, partition the two-dimensional space into four bins, A, B, C, and D, as illustrated in Fig.~\ref{fig:ABCDplane}.
The bin boundaries are optimized for the best expected search sensitivity, separately for each of the search categories.
Since the HNL is primarily produced back-to-back with the prompt lepton, the bin with the best signal-to-background ratio is bin D defined as $\dphilep > 2.8$ and $\nhits > 150$ or 200.
Similarly, bin C is defined as $\dphilep > 2.8$ and $\nhits \leq 150$ or 200, bin A is defined as $\dphilep < 2.8$ and $\nhits > 150$ or 200, and bin B containing the least signal is defined as $\dphilep < 2.8$ and $\nhits \leq 150$ or 200.
The responses of the DT and CSC detectors to shower particles are generally different with CSC signal clusters having a larger hit multiplicity compared with DT signal clusters.
The optimal bin boundary is found to be higher (200) for the CSC signal regions (SRs) compared with the DT SRs (150), with a similar signal efficiency of about 0.01\% for an HNL with a \ctau of 1\unit{m}.
Because of the independence of the two variables, the expected background event rate in the signal-enriched bin D can be related to the other three bins by $\lambda_\mathrm{D} = (\lambda_\mathrm{A}\lambda_\mathrm{C})/\lambda_\mathrm{B}$, where $\lambda_\mathrm{X}$ is the expected background event rate (i.e. the Poisson mean) in bin X.
To account for a potential signal contribution to bins A, B, and C, a binned maximum likelihood fit is performed simultaneously in the four bins, with a common signal strength parameter scaling the signal yields in each bin.
The background component of the fit is constrained to obey the ABCD relationship.

Because of the different background composition, we separate events with a selected prompt electron and a selected prompt muon into two disjoint categories.
We also separate events with clusters in the CSC and DT detectors.
Finally, for events in the category with a prompt muon and an MDS cluster in the DT detectors, we separate events into subcategories, which we call the DT-MB2 and DT-MB3/MB4, depending on whether the majority of the cluster's hits fall in the MB2 station or the MB3 or MB4 station of the DT detector.
This categorization is motivated by the fact that the background differs for clusters in the MB2 station because of the thinner shielding in front of it.
The additional subcategories for DT clusters are advantageous for the prompt muon channel because of its ability to suppress the additional \zmm background as detailed below.
This results in a total of five SRs.

The background estimation procedure is validated using events in the early OOT validation region (VR), defined as events passing all analysis selections except those related to the cluster time; instead, a negative cluster time is required.
Additionally, the background estimation procedure is also checked for the in-time VR, defined as events with the azimuthal angle between the MDS cluster centroid position and the \ptvecmiss: $\Delta\phi(\text{cluster},\ptvecmiss)>0.7$.
This VR is signal depleted because the MDS cluster and \ptvecmiss tend to align, as described in Section~\ref{sec:clusterSelection}.
The observed and predicted yields for these VRs are summarized in Table~\ref{tab:ABCDValidation} and show that the ABCD method results are consistent with observed background yields.
Similar consistencies are observed when the validations are performed in both in-time and OOT VRs with looser $\nhits$ requirements than in the SRs.

\begin{table*}[!htb]
  \centering
  \topcaption{
    Validation of the ABCD method in the OOT and in-time validation regions.
    The predictions of the method for the signal bin (last column) are consistent with the observed number of events, shown in the second-to-last column.
  }
  \label{tab:ABCDValidation}
  \begin{scotch}{lcccccc}
    Event category   & Validation region & A   & B     & C    & D & D (pred.)      \\
    \hline
    Muon, DT-MB2     & OOT               & 9   & 6924  & 944  & 0 & $1.2 \pm 0.4$  \\
    Muon, DT-MB3/MB4 & OOT               & 11  & 593   & 86   & 1 & $1.6 \pm 0.5$  \\
    Muon, CSC        & OOT               & 103 & 31074 & 4044 & 9 & $13.4 \pm 1.3$ \\
    Electron, DT     & OOT               & 14  & 3301  & 366  & 2 & $1.6 \pm 0.4$  \\
    Electron, CSC    & OOT               & 33  & 13774 & 1647 & 2 & $4.0 \pm 0.7$  \\[\cmsTabSkip]
    Muon, DT-MB2     & In time           & 10  & 5087  & 467  & 2 & $0.9 \pm 0.3$  \\
    Muon, DT-MB3/MB4 & In time           & 9   & 785   & 107  & 2 & $1.2 \pm 0.4$  \\
    Muon, CSC        & In time           & 31  & 7445  & 532  & 1 & $2.2 \pm 0.4$  \\
    Electron, DT     & In time           & 8   & 2446  & 220  & 0 & $0.7 \pm 0.3$  \\
    Electron, CSC    & In time           & 7   & 3217  & 227  & 0 & $0.5 \pm 0.2$
  \end{scotch}
\end{table*}
The ABCD background predictions for bin D of the SRs are listed in Table~\ref{tab:ABCDBkgSummary} for each of the event categories considered in this search.
The event yields for the bins A, B, and C are shown, as well as the prediction for the background in the signal-enriched bin D.

\begin{table}[!htb]
  \centering
  \topcaption{
    The event yields in the bins A, B, and C are shown in each of the event categories considered in the search, as well as the prefit prediction for the ABCD background in the signal-enriched bin D.
  }
  \label{tab:ABCDBkgSummary}
  \begin{scotch}{lccccc}
    Channel  & Region     & A  & B    & C    & D (pred.)     \\
    \hline
    Muon     & CSC        & 40 & 9316 & 1219 & $5.2 \pm 0.8$ \\
    Muon     & DT-MB2     & 16 & 6512 & 1048 & $2.6 \pm 0.7$ \\
    Muon     & DT-MB3/MB4 & 11 & 1011 & 229  & $2.5 \pm 0.8$ \\[\cmsTabSkip]
    Electron & CSC        & 10 & 3993 & 497  & $1.2 \pm 0.4$ \\
    Electron & DT         & 15 & 3069 & 463  & $2.3 \pm 0.6$
  \end{scotch}
\end{table}

In the prompt-muon event categories, \zmm events constitute a significant background source that is not predicted by the ABCD method.
The muon veto normally suppresses such backgrounds, but in rare cases, the muon may not be reconstructed in the muon system due to instrumental effects such as gaps between chambers.
These nonreconstructed muons will lead to \ptmiss in the same direction as the cluster, and thus these events are not present in the VRs listed in Table~\ref{tab:ABCDValidation}.

We define a \zmm control region (CR) by inverting the MB1, or ME1/1 or ME1/2 hit veto requirements for MDS clusters in the DT or CSC detectors, respectively.
After that, bin D will be dominated by the \zmm background.
The \zmm expected background yield in bin D of the \zmm CR is calculated as $\lambda_{\zmm,\mathrm{D}}^{\mathrm{CR}} = N_{\mathrm{D}}^{\mathrm{CR}} -  \lambda_{\mathrm{ABCD~bkg,D}}^{\mathrm{CR}}$, where $N_{\mathrm{D}}^{\mathrm{CR}}$ is the observed data yield in bin D and $\lambda_{\mathrm{ABCD~bkg,D}}^{\mathrm{CR}}$ is the ABCD method prediction.
We extrapolate $\lambda_{\zmm,\mathrm{D}}^{\mathrm{CR}}$ to the SR by applying a transfer factor $\zeta$, which estimates the pass-to-fail efficiency ratio for the MB1, ME1/1, and ME1/2 veto requirements for MDS clusters produced by muon bremsstrahlung.

  {\tolerance=2000
    The factor $\zeta$ is measured in a sample enhanced in MDS clusters produced by muon bremsstrahlung obtained by selecting dileptonic decays of \ttbar pairs with an electron and muon in the final state.
    We select events with: (i) one prompt electron matched to an electron trigger object; (ii) no reconstructed muon; (iii) one MDS cluster passing all the veto selections except the MB1, ME1/1, or ME1/2 vetoes; and (iv) exactly two additional jets with $\pt>20\GeV$, $\abs{\eta} < 2.4$ that pass the medium working point of the combined secondary vertex (CSV) b-tagging algorithm~\cite{btag,BTV-16-002}, which has an efficiency of 60\% and a mistag rate of 1\% for light-flavor or gluon jets.
    We also require that the MDS cluster is geometrically separated by $\Delta R>0.8$ from the two jets that satisfy the CSV algorithm requirement to ensure that the MDS cluster is not produced by a punch-through jet.
    These requirements result in an event sample that is pure in dileptonic \ttbar events, in which both the electron and jets cannot produce the MDS cluster, thus ensuring that the MDS cluster is produced by the unreconstructed muon.
    We measure the $\zeta$ factors separately for MDS clusters in the DT and CSC detectors.
    We observed a linear dependence of $\zeta$ on \nhits in the MB2 category.
    To account for this dependence, we perform a linear fit to the data in the CR and evaluate the fitted function at $\nhits=150$ as the $\zeta$ for the MB2 category.
    Finally, the expected \zmm background contribution to bin D of the SR is calculated as $\lambda_{\zmm,\mathrm{D}}^{\mathrm{SR}} = \zeta \lambda_{\zmm,\mathrm{D}}^{\mathrm{CR}}$.
    Details of the \zmm background prediction are summarized in Table~\ref{tab:ZmumuBkgEstimate}.
    \par}

To validate the \zmm background estimation method, we define another VR as a subset of the MB2 SR with $\nhits \leq 120$ which is expected to have a negligible signal contribution.
In this VR, we define the bin boundaries for the ABCD method to be at 2.8 in \dphilep and at 110 in \nhits.
We perform the prediction of the \zmm and non-muon-induced background events using the same method applied to the full SR and obtain predictions of $2.7\pm1.6$ for the \zmm background and $9.9\pm1.3$ for the other background, for a total background prediction of $12.6 \pm 2.1$ events.
We observe an event yield of 12, which is consistent with our background predictions.

\begin{table*}[!htb]
  \centering
  \topcaption{
    Summary of the \zmm background estimate in different categories.
    The first three columns show the estimates in the \zmm enriched control region of the total background and its \zmm and non-muon-induced components.
    The fourth column shows the transfer factors $\zeta$ used to predict the \zmm background in the signal region, shown in the fifth column.
  }
  \label{tab:ZmumuBkgEstimate}
  \begin{scotch}{lccccc}
    Region     & $N_D^\mathrm{CR}$ & $\lambda_{\mathrm{ABCD~bkg,D}}^{\mathrm{CR}}$ & $\lambda_{\zmm,\mathrm{D}}^{\mathrm{CR}}$ & $\zeta$         & $\lambda_{\zmm,\mathrm{D}}^{\mathrm{SR}}$ \\
    \hline
    CSC        & 129               & $45 \pm 2$                                    & $84 \pm 12$                               & ($4.8\pm1.3$)\% & $3.9 \pm 1.2$                             \\
    DT-MB2     & 35                & $12.2 \pm 1.5$                                & $22.8 \pm 6.1$                            & ($36\pm31$)\%   & $8.2 \pm 7.4$                             \\
    DT-MB3/MB4 & 6                 & $2.9 \pm 0.7$                                 & $3.1 \pm 2.6$                             & ($2\pm1$)\%     & $0.06 \pm 0.06$
  \end{scotch}
\end{table*}

\section{Systematic uncertainties}
\label{sec:systematics}

The dominant systematic uncertainties in this search are those in the background predictions.
The main source of uncertainty for the ABCD method arises from the statistical uncertainties of the background-enriched bins A, B, and C.
This uncertainty accounts for 27\% and 22\% (40\% and 18\%) of the size of the total background in the electron and muon categories, respectively, for MDS clusters in the DT (CSC) detector.

For the prompt-muon categories, an uncertainty in the estimate of the \zmm background also contributes significantly to the total uncertainty.
The uncertainty for that background prediction arises from the statistical uncertainty of the $\zeta$ measurement, accounting for about 72\%, 2\%, and 15\% of the size of the total background in the muon channel for DT-MB2, DT-MB3/MB4, and CSC clusters, respectively.

Systematic uncertainties in the signal yield include both theoretical and instrumentation effects.
The theoretical uncertainty in the inclusive cross section of \PW boson production is 3.8\%, which is dominated by the PDF uncertainty.
The uncertainty in the \PW boson \pt distribution is estimated by varying the renormalization and factorization scales by a factor of two separately and coherently, and evaluating the size of the envelope in the resulting signal yield, which is found to be 1.6\%.
The uncertainty of parton shower modeling is estimated similarly by varying the renormalization scales for parton showers and is found to be 4\%.

\begin{figure*}[!htb]
  \centering
    \includegraphics[width=0.45\textwidth]{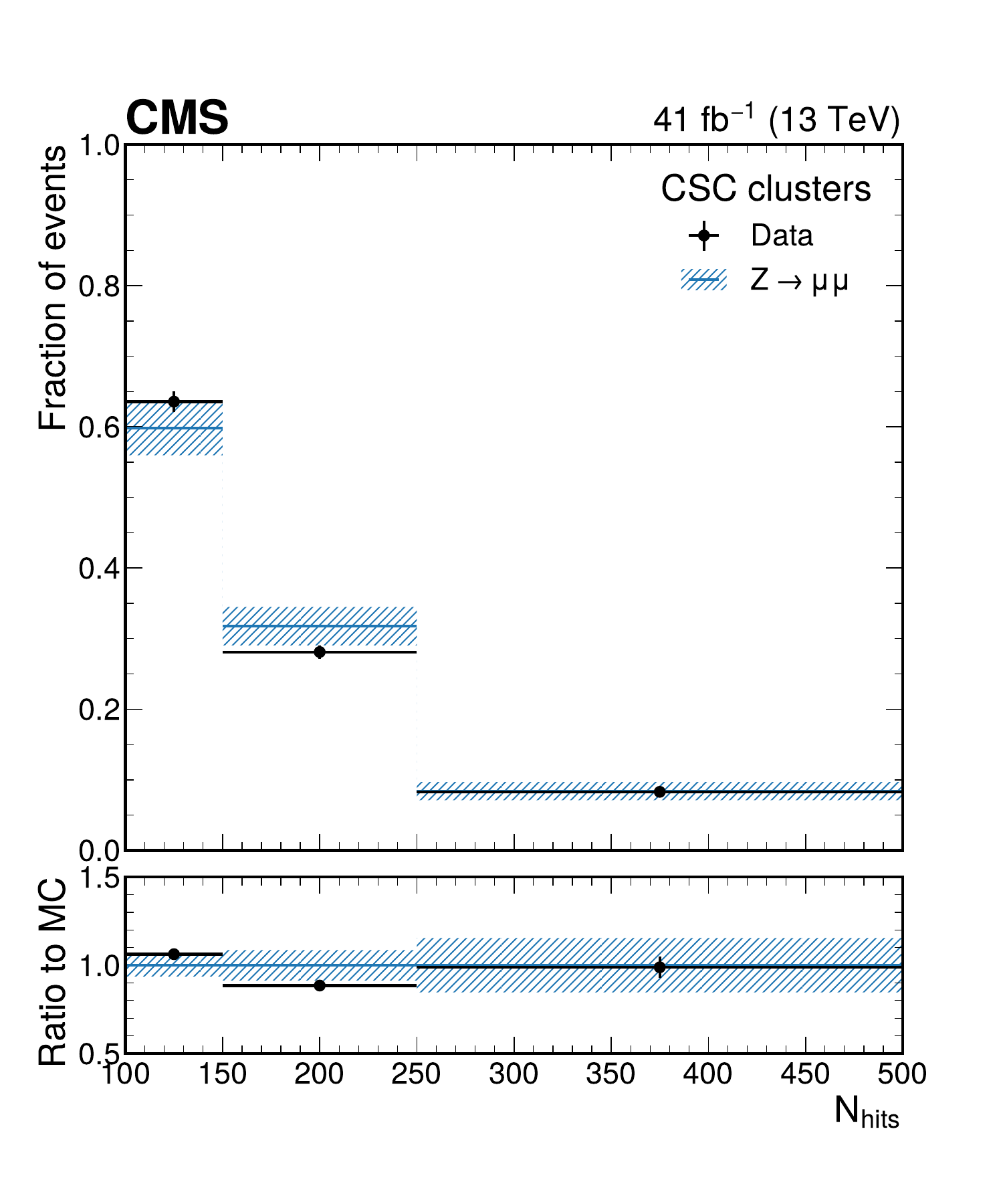}
    \includegraphics[width=0.45\textwidth]{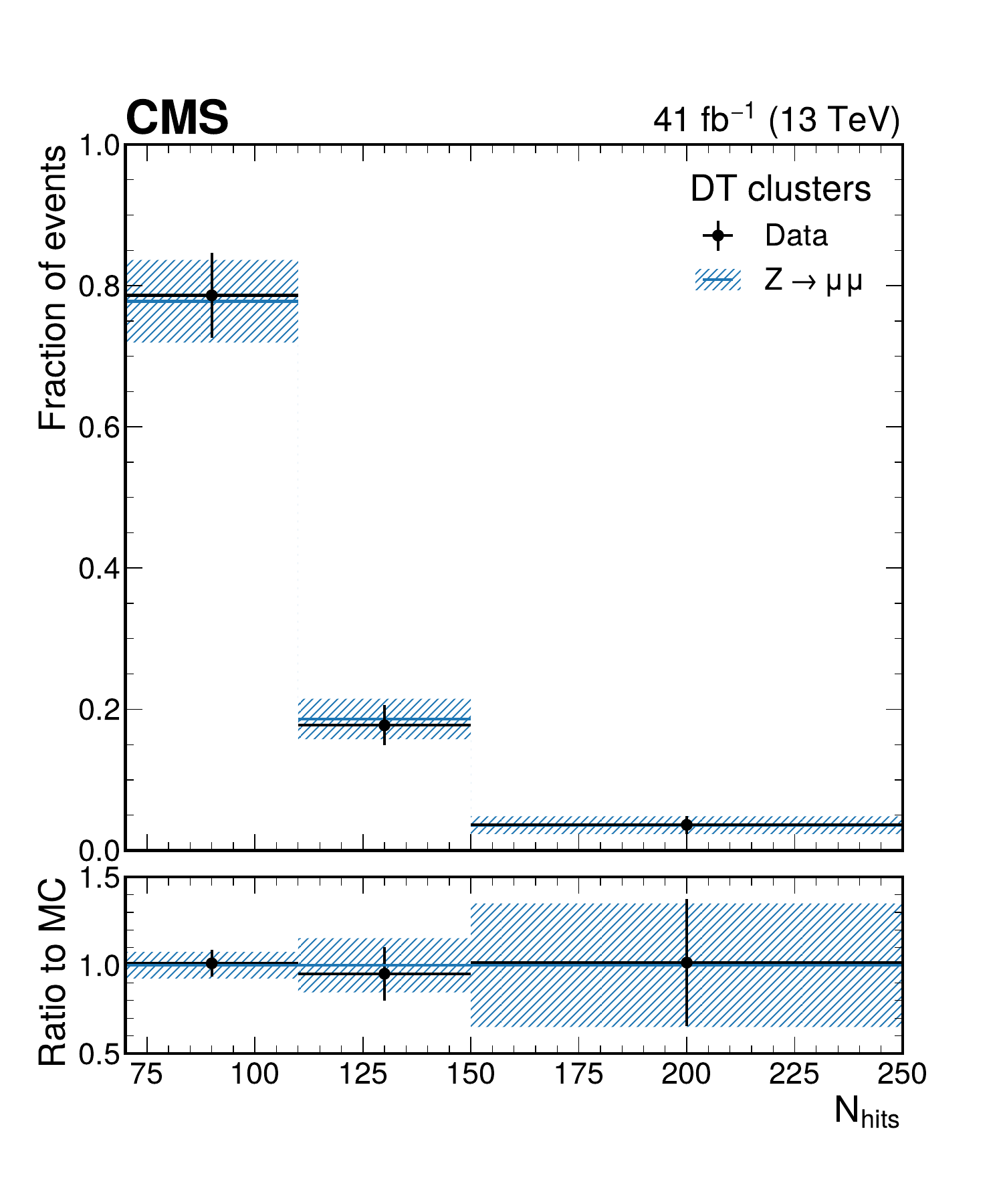}
    \caption{Comparison of \nhits distributions for events with muons from \zmm between data and simulation, for CSC clusters (left) and DT clusters (right), using data collected in 2017 and the simulation of the corresponding data-taking conditions. The data sample is selected by requiring a two-muon invariant mass consistent with a Z boson and one of the muons is matched to an MDS cluster. Data-to-simulation correction factors are applied to the \zmm simulation. Only statistical uncertainties are included in the figure. Distributions made using data collected in 2016 and 2018 are found to have similar shape as the 2017 data sample.}
  \label{fig:clusterSim}
\end{figure*}

The accuracy of the CMS simulation and the GEANT4 software in describing the details of the evolution of electromagnetic and hadronic showers has already been validated extensively in past measurements.
However, the accuracy of the CMS simulation implementation of the response of the muon detectors in an environment with a large multiplicity of secondary particles, which affects the simulated reconstruction efficiencies in the SR, has not been explicitly verified.
This aspect is validated by comparing clusters produced in \zmm data and simulated events, in which one of the muons undergoes bremsstrahlung in the muon detectors and the associated photon produces an electromagnetic shower.
With this comparison, we derive the uncertainties in the cluster reconstruction efficiency for both CSC and DT clusters in simulation; they account for 16\% and 13\% of the signal yield for the DT and CSC categories, respectively.
These uncertainties apply to both electromagnetic and hadronic showers.
Figure~\ref{fig:clusterSim} illustrates this validation of the cluster simulation.
Data-to-simulation corrections on other cluster properties are also derived with the same method, and the uncertainties in the corrections are propagated as systematic uncertainties.
For DT clusters, a correction of 6.8\% is applied for the MB1 veto efficiency, with an uncertainty of 7.4\%.
For CSC clusters, corrections are applied to account for the hit and segment vetoes (2.8\%), muon veto (6.8\%), and jet veto efficiencies (2.1\%), with uncertainties of 0.1\%, 4.5\%, and 0.06\%, respectively.
Additionally, the uncertainties on efficiencies of the CSC cluster identification, time, and time spread requirements are estimated to be 5\%, 0.9\%, and 2.8\%, respectively.
In data, only the reconstructed hits that have at least two cathode hits in different CSC layers and match a given predefined pattern are read out.
In contrast, this readout condition is assumed to be satisfied in the signal simulation, which could lead to an overestimation of the \nhits in signal clusters.
We estimated this effect by excluding those chambers with less than 6 hits in signal clusters, and assigning the change in signal efficiency (1\%) as an uncertainty.

We also propagate additional systematic uncertainties that have a minor impact on the signal yield prediction.
They include uncertainties due to pileup, integrated luminosity, jet energy scale, and prompt electron or muon trigger and selection efficiencies.
The integrated luminosities for the 2016, 2017, and 2018 data-taking years have 1.2--2.5\% individual uncertainties~\cite{CMS-LUM-17-003,CMS-PAS-LUM-17-004,CMS-PAS-LUM-18-002}, while the overall uncertainty for the 2016--2018 period is 1.6\%.

Finally, the uncertainties on the signal yields from the limited numbers of simulated events are in the ranges 5--10\% depending on \mhnl and lifetime.
Theoretical and experimental uncertainties that are not related to the clusters are treated as fully correlated across different event categories.
Experimental uncertainties related to the clusters are treated as fully uncorrelated.
A full list of systematic uncertainties affecting the predicted signal yield is shown in Table~\ref{tab:signalSystematics}.

\begin{table}[!ht]
  \centering
  \topcaption{
    Summary of systematic uncertainties affecting the signal yield prediction.
    For DT clusters, the systematic uncertainties due to jet and muon vetoes are found to be negligible and are omitted.
    The uncertainties are reported relative to their impact on the predicted signal yield.
    \label{tab:signalSystematics}
  }
  \begin{scotch}{lcc}
    Systematic uncertainty  & Object      & Size of unc. [\%] \\
    \hline
    Integrated luminosity   & \NA         & 1.6               \\
    Pileup                  & \NA         & 1                 \\
    \PW boson cross section & \NA         & 3.8               \\
    \PW boson \pt           & \NA         & 1.6               \\
    Parton shower modeling  & \NA         & 4               \\[\cmsTabSkip]
    Trigger                 & Muon        & $<$0.1            \\
    Identification          & Muon        & 0.4--0.5          \\
    Isolation               & Muon        & 0.2--0.6          \\
    Trigger                 & Electron    & 0.2--0.3          \\
    Identification          & Electron    & 2.2--8.0          \\
    Jet energy scale        & \ptmiss     & 2.0               \\[\cmsTabSkip]
    Cluster reconstruction  & CSC cluster & 13                \\
    Cutoff-based ID         & CSC cluster & 5.1               \\
    Jet veto                & CSC cluster & 0.06              \\
    Muon veto               & CSC cluster & 4.5               \\
    CSC readout             & CSC cluster & 1.0               \\
    Hits and segment veto   & CSC cluster & 0.1               \\
    Cluster time            & CSC cluster & 0.9               \\
    Cluster time spread     & CSC cluster & 2.8               \\[\cmsTabSkip]
    Cluster reconstruction  & DT cluster  & 16                \\
    MB1 veto                & DT cluster  & 7.4
  \end{scotch}
\end{table}

\section{Results and interpretation}
\label{sec:results}
\begin{figure}[!htb]
  \begin{center}
    \includegraphics[width=\cmsFigWidth]{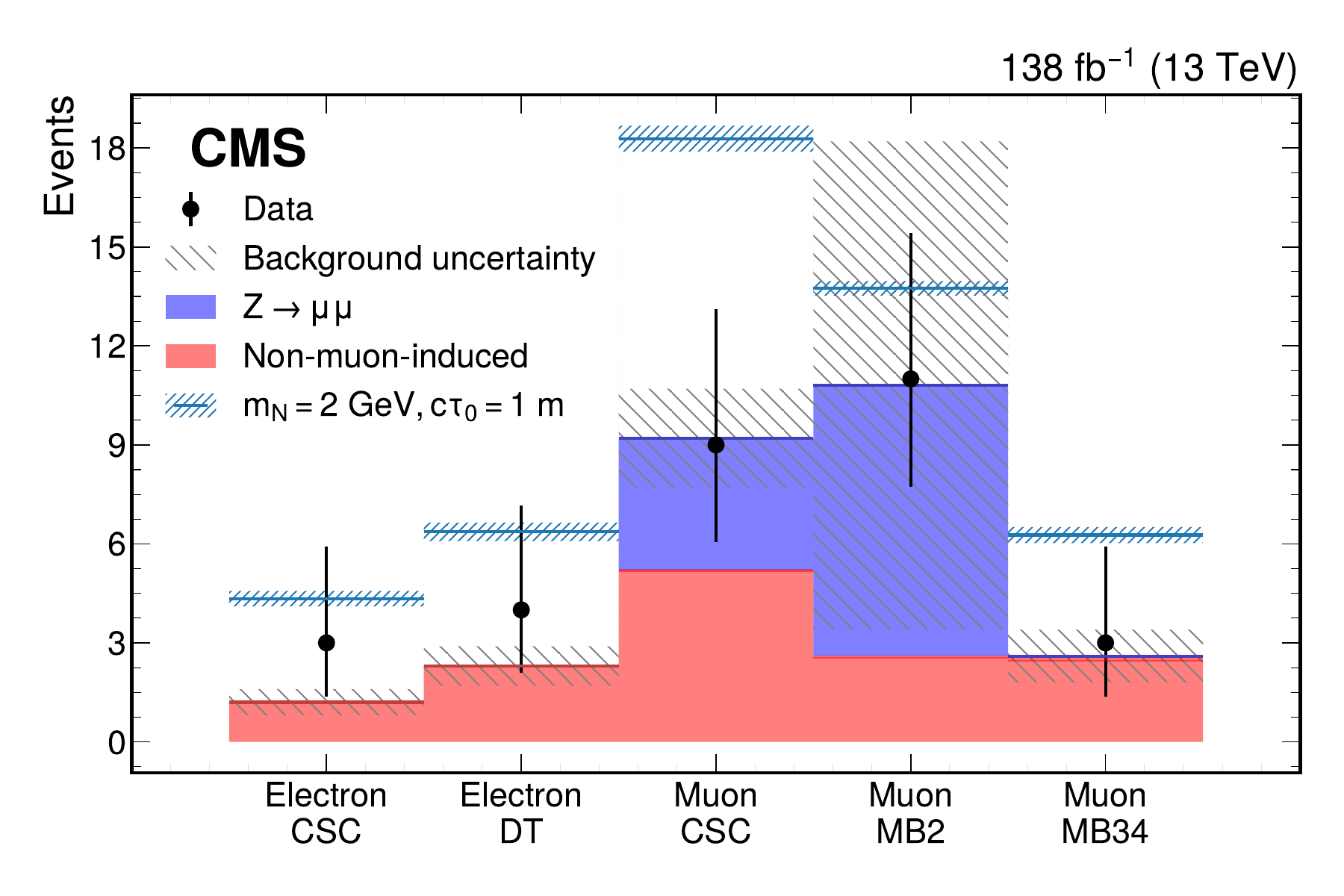}
    \caption{
      The expected and observed number of events in the signal region (bin D) of different event categories.
      Signal yields of a 2\GeV Majorana HNL with the mean proper decay length of 1\unit{m} are added to the expected background.
    }
    \label{fig:result}
  \end{center}
\end{figure}

Figure~\ref{fig:result} shows the expected and observed number of events in the SRs of the different event categories and the corresponding background predictions.
The observed yields agree with the predicted background in all channels.

No excess of data events over the background prediction is observed, and upper limits on the HNL production cross sections are evaluated using the \CLs criterion~\cite{JUNK1999435,Read_2002}, with the binned profile likelihood ratio~\cite{CMS-NOTE-2011-005} as the test statistic.
The likelihood is constructed as the product of Poisson distributions with mean values based on the predicted event rates across all five event categories.
The predicted event rates in each bin include the background yields predicted with the ABCD method plus the \zmm background contribution for bin D and the signal yields obtained from simulated events.
Systematic uncertainties in the predicted event rates are incorporated in the likelihood as nuisance parameters with log-normal constraints.
The asymptotic formulas~\cite{Cowan:2010js} are used to evaluate the exclusion limits for each HNL signal scenario.
The exclusion limits derived from the asymptotic formulas are consistent within 10\% with those based on pseudo-experiments.

Due to the lower \pt thresholds for the prompt muons, the signal yields in the muon channel are larger than those of the electron channels for the same \mhnl and lifetime.
However, the presence of the \zmm background reduces the overall sensitivity of the muon channel to a level comparable to the electron channel.
For both the muon and electron channels, the CSC category contributes the majority of the overall sensitivity.
This is because the CSC category has smaller background rates compared with the DT category, hence retaining more signal events at the optimal thresholds for background rejection.

\begin{figure*}[!htb]
  \centering
  \includegraphics[width=0.4\textwidth]{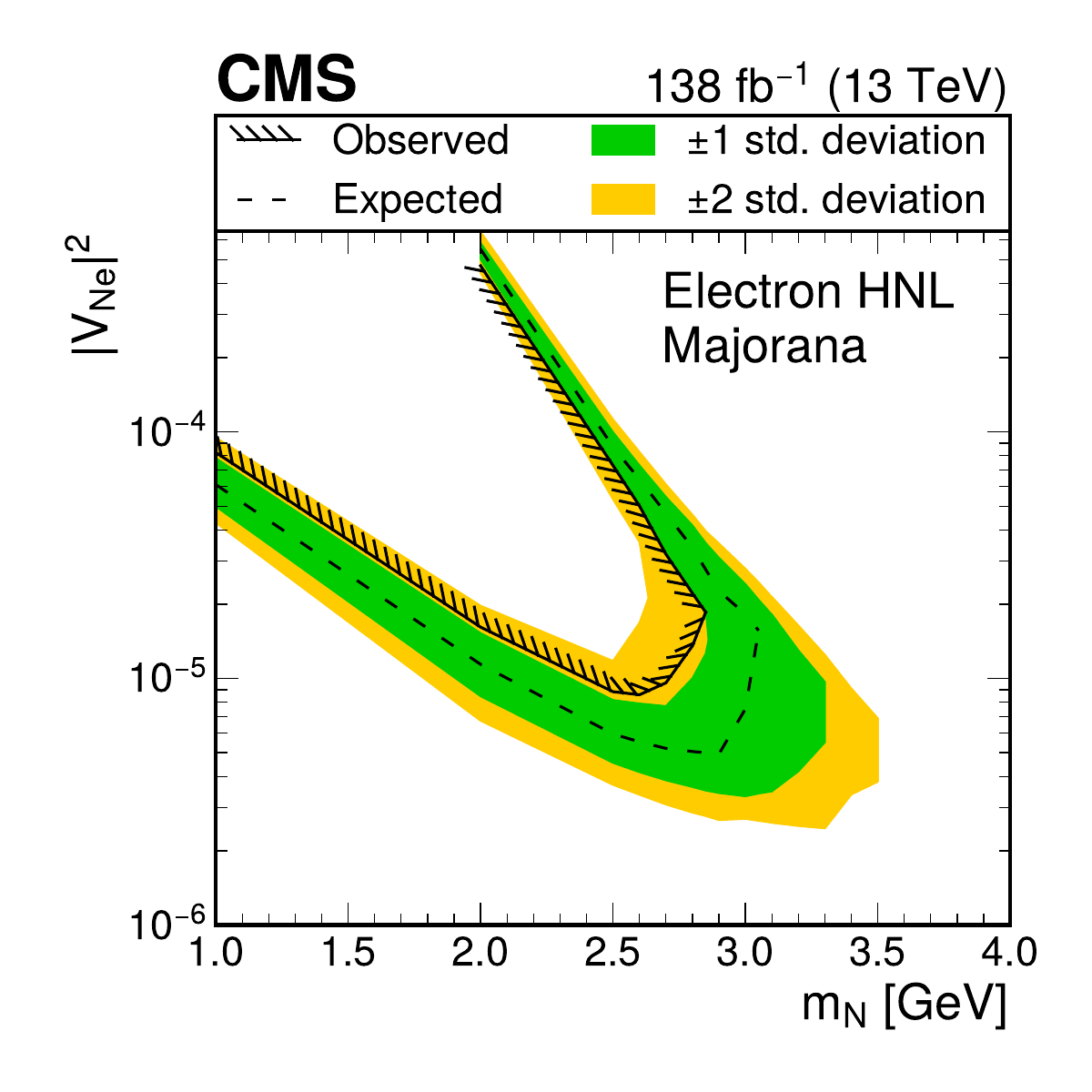}
  \includegraphics[width=0.4\textwidth]{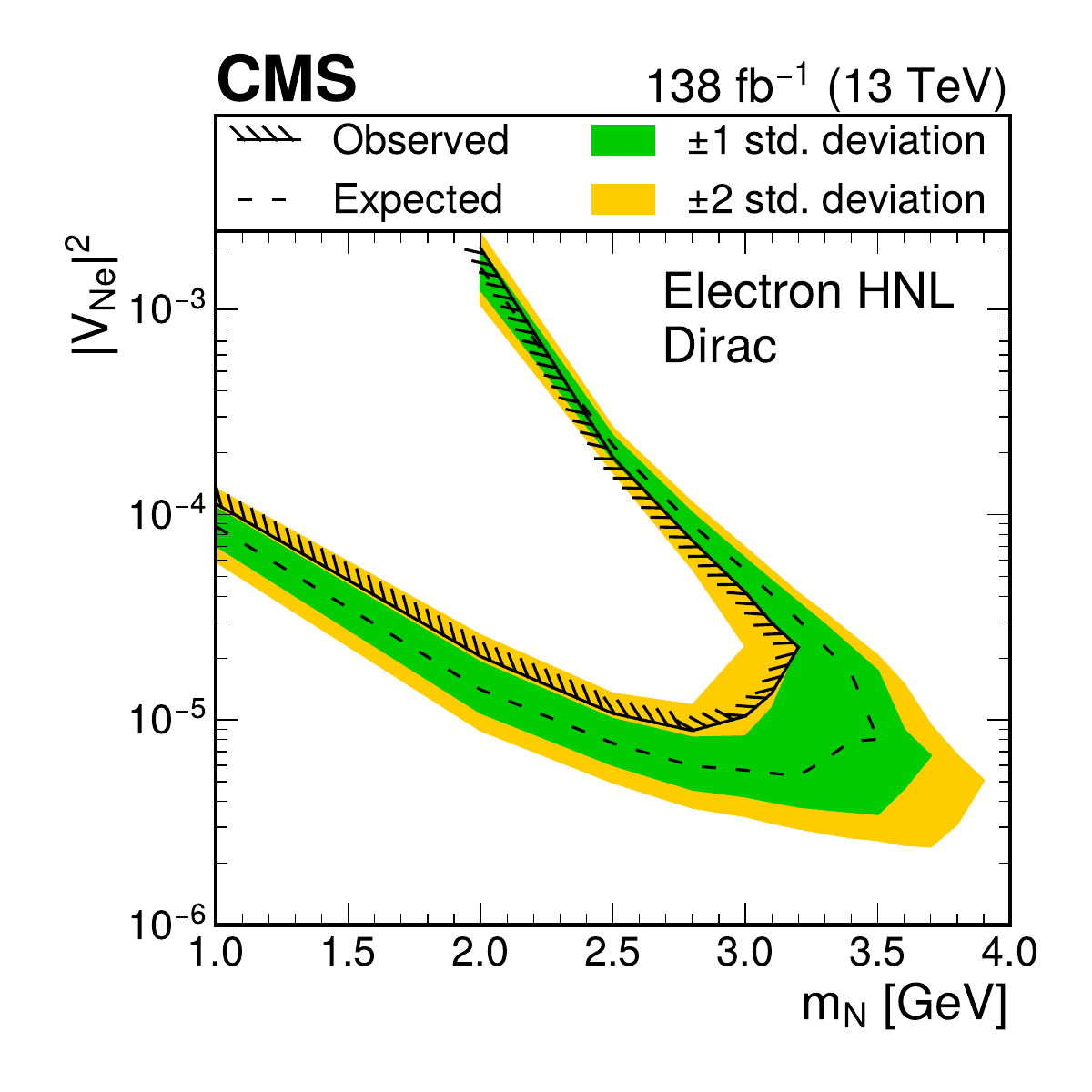}
  \includegraphics[width=0.4\textwidth]{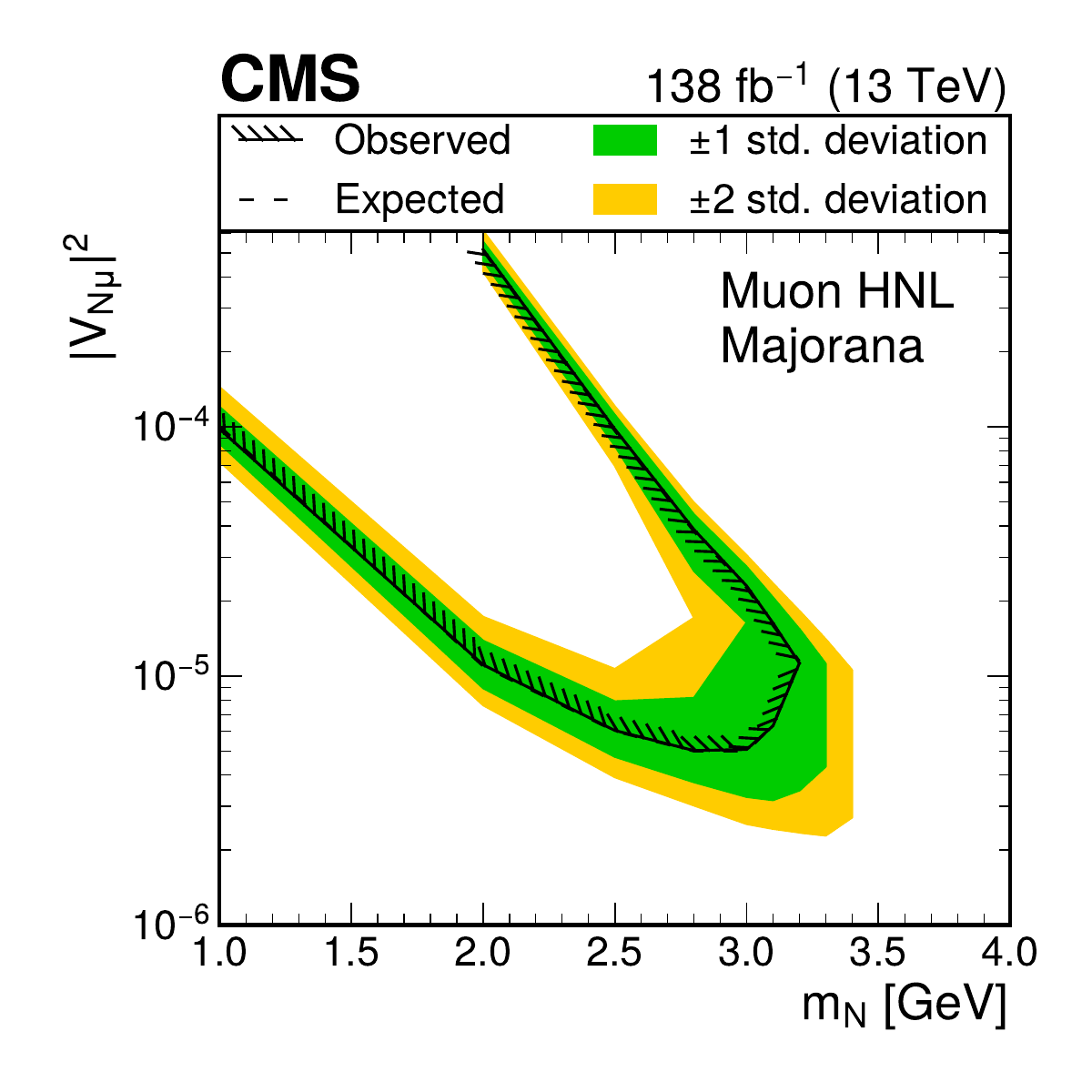}
  \includegraphics[width=0.4\textwidth]{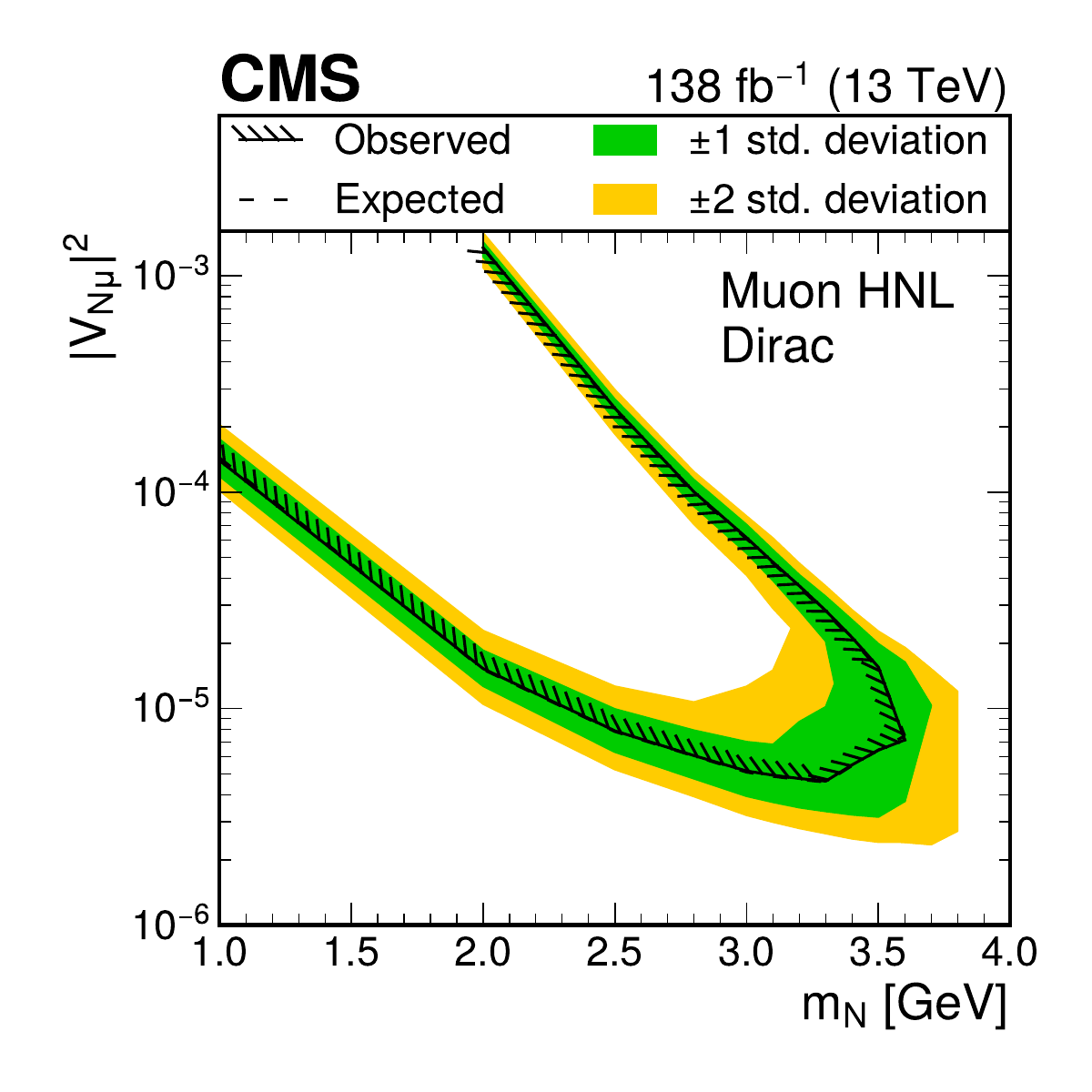}
  \includegraphics[width=0.4\textwidth]{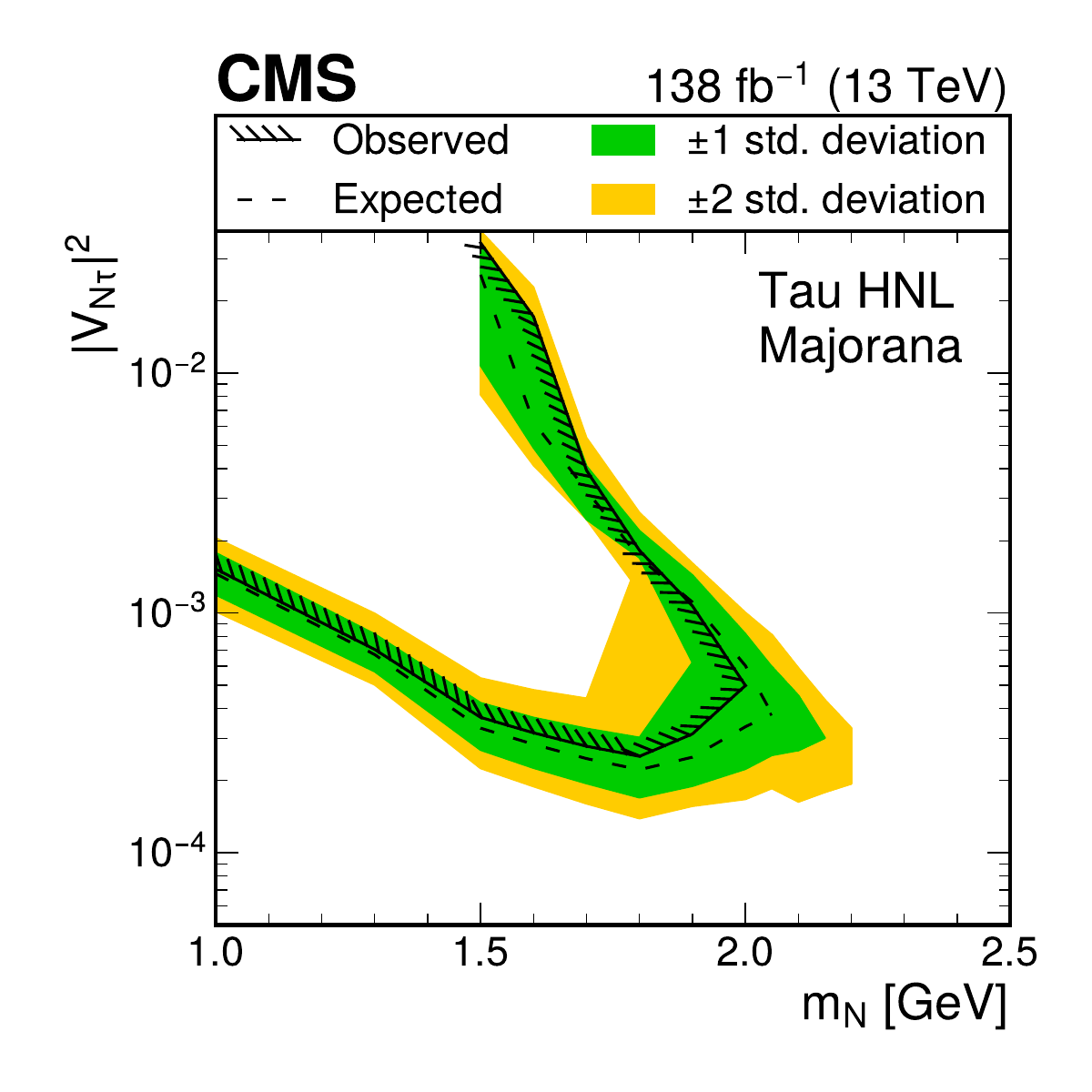}
  \includegraphics[width=0.4\textwidth]{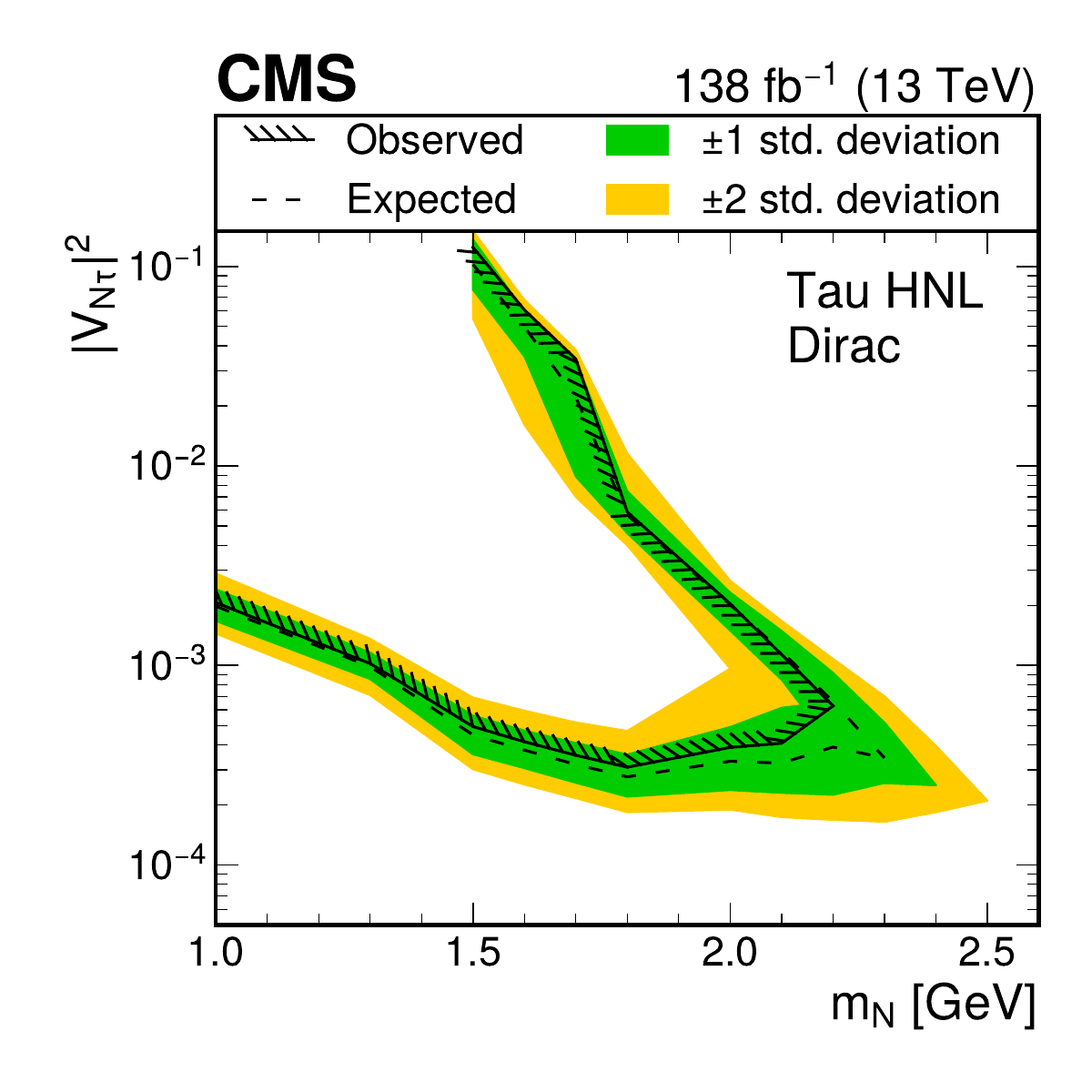}
  \caption{
    Expected and observed upper 95\% \CL limits on \VNesq (upper), \VNmusq (middle) and \VNtausq (lower) as functions of the HNL mass~(\mhnl) for a Majorana (left) and Dirac (right) type HNL.
    The $\PGt$ neutrino mixing limit is obtained by combining the results from the electron and muon channels.
    For these limit calculations, the HNL is assumed to mix with a single lepton flavor state only.
    The differences between the expected and observed limits on \VNmusq are not visible in this figure.
  }
  \label{fig:AsymLimit_V2_majorana}
\end{figure*}

Figure~\ref{fig:AsymLimit_V2_majorana} shows the expected and observed upper limits at 95\% confidence level (\CL) on the HNL mixing parameters \VNesq, \VNmusq, and \VNtausq, as functions of \mhnl for the Majorana and Dirac HNL interpretations.
For \VNesq (\VNmusq), only events passing the electron (muon) selections are used to evaluate the limits.
For \VNtausq, both the events passing the electron and muon selections are used, because the prompt $\PGt$ lepton can decay into an electron or muon.
The limits of \VNlsq versus \mhnl feature an upper branch (short lifetime) and a lower branch (long lifetime).
The lower acceptance in the short lifetime branch is compensated by an increased cross section, and thus achieves a signal yield comparable with the long lifetime branch.
Below \mhnl of 2.0 (1.5)\GeV for electron or muon ($\PGt$) type HNLs, we do not calculate a limit in the short lifetime branch because the signal acceptance in the muon system approaches zero and renders the calculation inaccurate.
Furthermore, in that region other experimental results~\cite{BEBC,CHARM:1985nku,NuTeV} place more stringent limits.
Table~\ref{tab:result_summary} summarizes the observed limits on \VNlsq for Majorana and Dirac type HNL of this search.
This result sets the most stringent limits to date in \VNlsq for HNL masses in the range 2.1--3.0 (1.9--3.3)\GeV for electron (muon) neutrino mixing parameters~\cite{hepdata}.
Above 3.0 (3.3)\GeV, previous CMS results~\cite{CMS:2022fut} place tighter limits on electron (muon) mixing parameters.

\begin{table*}[!htb]
  \centering
  \topcaption{
    Excluded ranges of \VNlsq for Majorana and Dirac type HNLs at select HNL masses.
    The chosen HNL masses are those at which the excluded values of \VNlsq have the smallest magnitude.
  }
  \label{tab:result_summary}
  \begin{scotch}{lccc}
    Mixing parameter & HNL type & \mhnl (\GeVns) & Excluded range at 95\% \CL             \\
    \hline
    \VNesq           & Majorana & 2.6            & $8.6\times10^{-6}$--$5.0\times10^{-5}$ \\
    \VNmusq          & Majorana & 2.8            & $5.0\times10^{-6}$--$3.9\times10^{-5}$ \\
    \VNtausq         & Majorana & 1.8            & $2.5\times10^{-4}$--$1.8\times10^{-3}$ \\[\cmsTabSkip]
    \VNesq           & Dirac    & 2.8            & $8.9\times10^{-6}$--$7.5\times10^{-5}$ \\
    \VNmusq          & Dirac    & 3.3            & $4.6\times10^{-6}$--$2.8\times10^{-5}$ \\
    \VNtausq         & Dirac    & 1.8            & $3.1\times10^{-4}$--$5.9\times10^{-3}$
  \end{scotch}
\end{table*}

\begin{figure*}[!htb]
  \centering
  \includegraphics[width=0.48\textwidth]{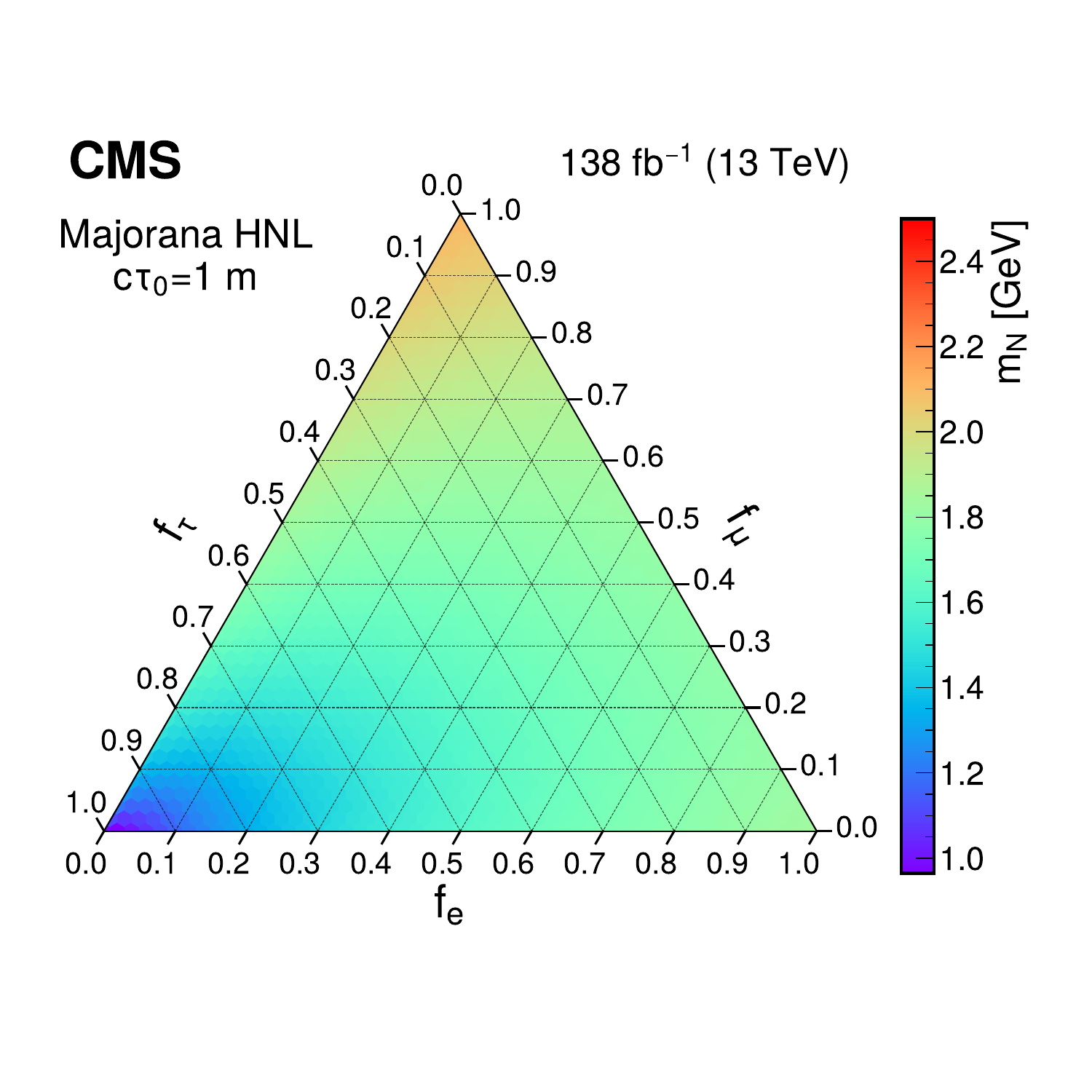}
  \includegraphics[width=0.48\textwidth]{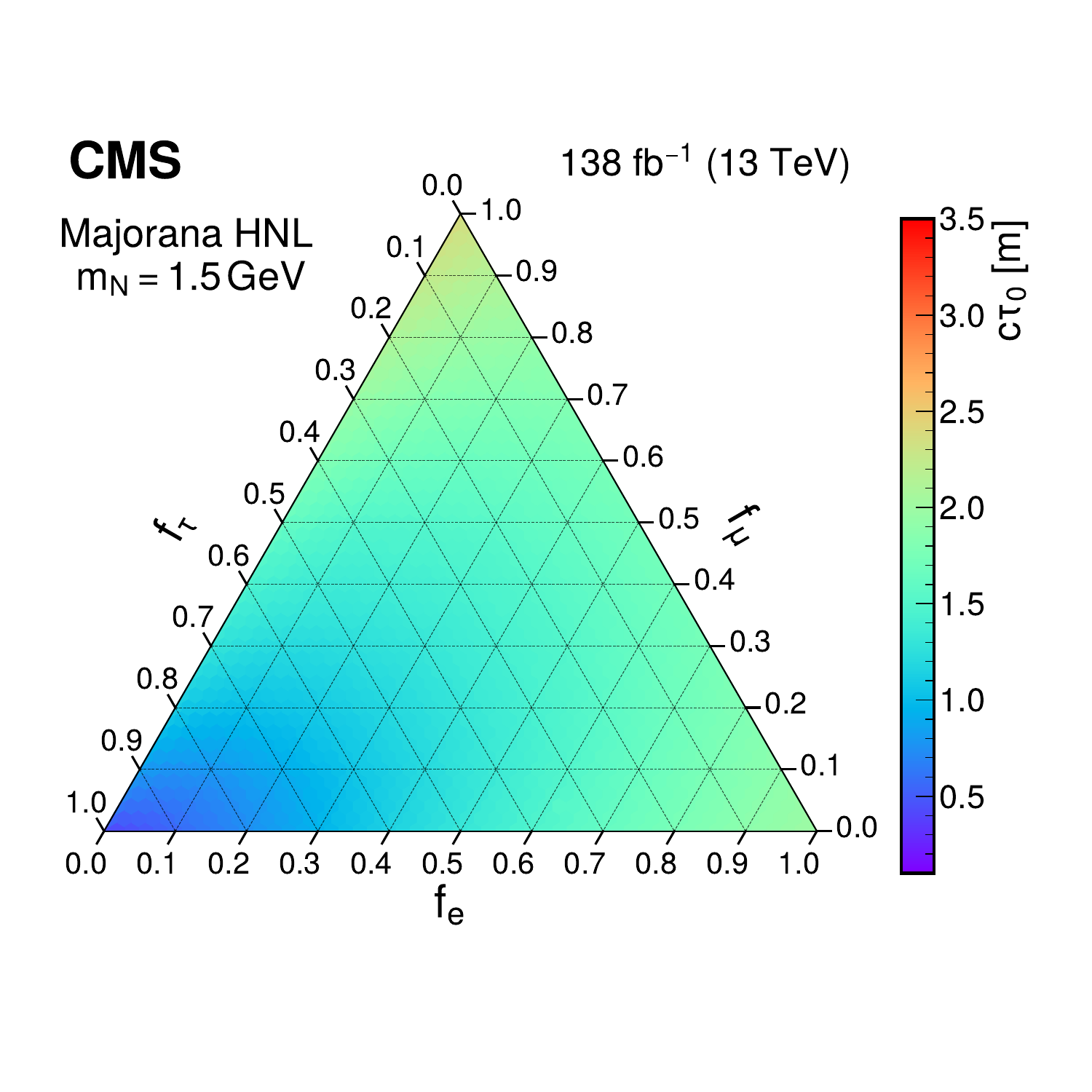}
  \includegraphics[width=0.48\textwidth]{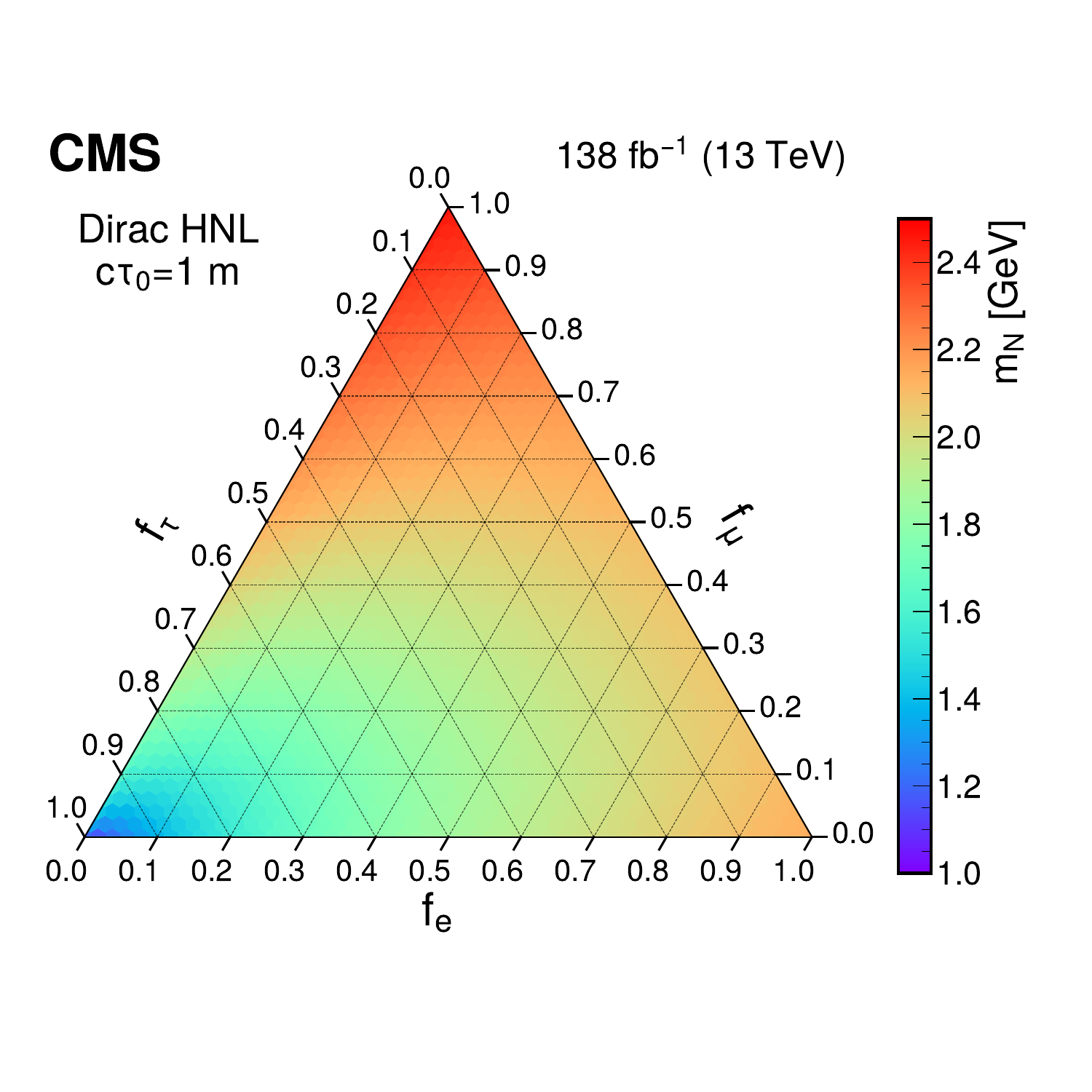}
  \includegraphics[width=0.48\textwidth]{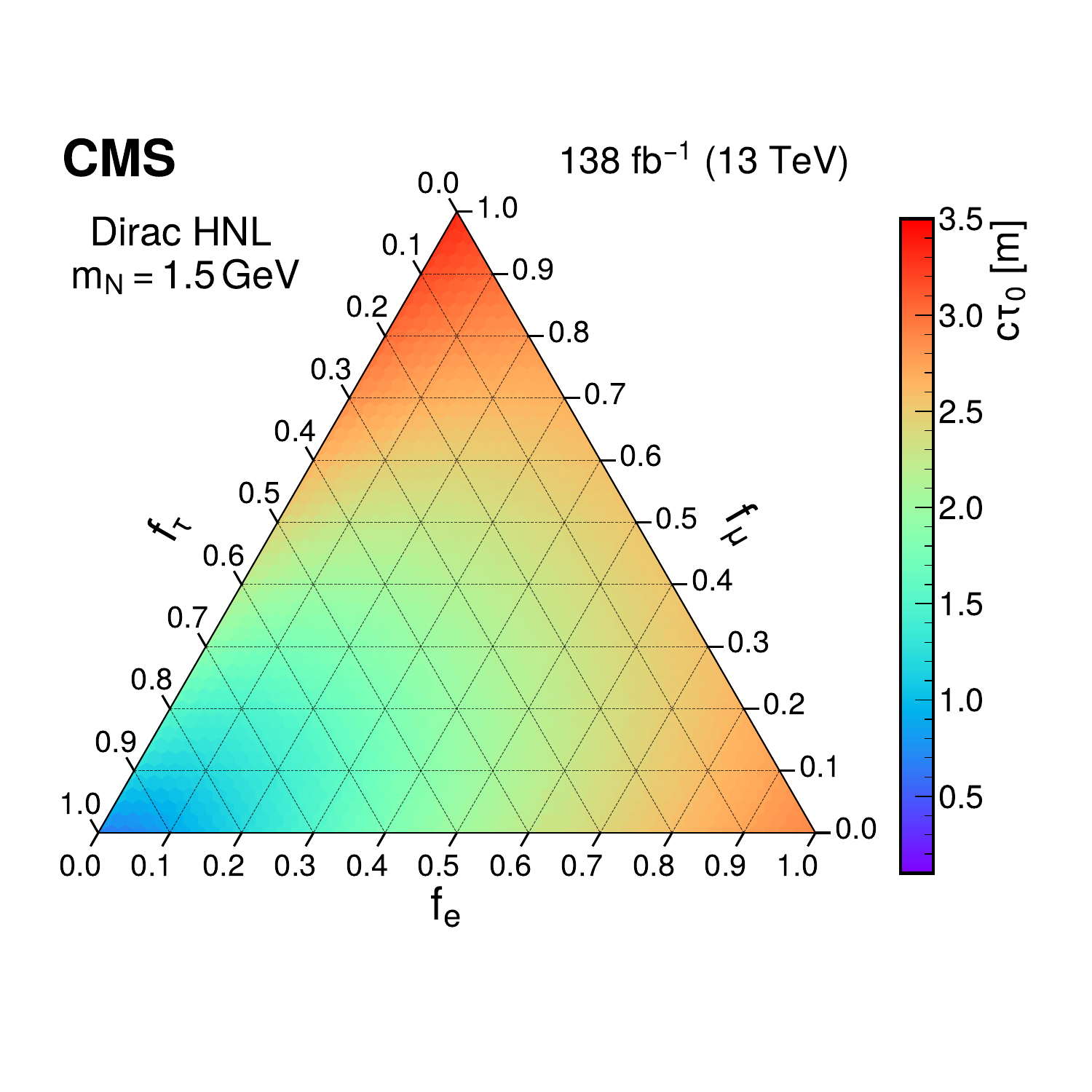}
  \caption{
    The largest values of the Majorana (upper) and Dirac (lower) HNL mass (left) and mean proper decay length (right) parameters that are excluded at 95\% \CL are shown as a function of the mixing matrix elements squared ratios $f_\ell$ with the three lepton generations, considering a mean proper decay length of 1\unit{m} and a fixed mass of 1.5\GeV, respectively.
  }
  \label{fig:Limit_triangle}
\end{figure*}

Signal yields for mixing among several different lepton flavors are obtained by reweighting the single flavor signal yields with the same \mhnl and \ctau.
In this case, the signal yields for multiple neutrino species mixing are proportional to \VNlsq.
The ratio of \VNlsq to the sum of the mixing matrix elements squared is defined as $f_\ell =\VNlsq/(\VNesq+\VNmusq+\VNtausq)$, and sums to unity by construction.
For a fixed value of \mhnl (\ctau), we scan the three $f_\ell$ parameters and exclude a range in the \ctau (\mhnl) parameter for each set of $f_\ell$ values.
On the left (right) of Fig.~\ref{fig:Limit_triangle}, we show the largest values of \mhnl (\ctau) that are excluded as a function of the $f_\ell$ parameters.
The sensitivity across the $f_\ell$ plane varies mainly because of the different trigger and reconstruction efficiencies of the prompt lepton.
The limits at $f_\Pe{:}f_\PGm{:}f_\PGt = 0{:}\frac{1}{2}{:}\frac{1}{2}$ and $\frac{1}{3}{:}\frac{1}{3}{:}\frac{1}{3}$ probe parameter space consistent with the constraints from neutrino oscillation data in the minimal seesaw scenarios~\cite{Drewes_2022}.

\section{Summary}
\label{sec:summary}

A search for long-lived Dirac or Majorana heavy neutral leptons (HNLs) has been performed using proton-proton collision data at $\sqrt{s} = 13\TeV$, corresponding to an integrated luminosity of 138\fbinv.
The search targets events with one prompt electron or muon and a muon detector shower (MDS) that would result from HNL decays occurring in the CMS muon detector.
The presence of the MDS signature along with the associated vetoes and identification criteria suppresses the standard model background by a factor exceeding $10^{7}$, while maintaining typical signal efficiencies of 25--35\%.
No significant excess over the standard model background is observed.
The results are interpreted as 95\% confidence level limits on the HNL mixing matrix elements squared \VNesq, \VNmusq, and \VNtausq.
We also present limits on the HNL mass and mean proper decay length as a function of the mixing matrix element squared fractions to the three lepton generations.
The most stringent limits to date for HNLs in the mass range of 2.1--3.0 (1.9--3.3)\GeV are set, reaching squared mixing matrix element values as low as $8.6\,(4.6) \times10^{-6}$ in the electron (muon) channel.

\begin{acknowledgments}
We congratulate our colleagues in the CERN accelerator departments for the excellent performance of the LHC and thank the technical and administrative staffs at CERN and at other CMS institutes for their contributions to the success of the CMS effort. In addition, we gratefully acknowledge the computing centers and personnel of the Worldwide LHC Computing Grid and other centers for delivering so effectively the computing infrastructure essential to our analyses. Finally, we acknowledge the enduring support for the construction and operation of the LHC, the CMS detector, and the supporting computing infrastructure provided by the following funding agencies: SC (Armenia), BMBWF and FWF (Austria); FNRS and FWO (Belgium); CNPq, CAPES, FAPERJ, FAPERGS, and FAPESP (Brazil); MES and BNSF (Bulgaria); CERN; CAS, MoST, and NSFC (China); MINCIENCIAS (Colombia); MSES and CSF (Croatia); RIF (Cyprus); SENESCYT (Ecuador); ERC PRG, RVTT3 and MoER TK202 (Estonia); Academy of Finland, MEC, and HIP (Finland); CEA and CNRS/IN2P3 (France); SRNSF (Georgia); BMBF, DFG, and HGF (Germany); GSRI (Greece); NKFIH (Hungary); DAE and DST (India); IPM (Iran); SFI (Ireland); INFN (Italy); MSIP and NRF (Republic of Korea); MES (Latvia); LMTLT (Lithuania); MOE and UM (Malaysia); BUAP, CINVESTAV, CONACYT, LNS, SEP, and UASLP-FAI (Mexico); MOS (Montenegro); MBIE (New Zealand); PAEC (Pakistan); MES and NSC (Poland); FCT (Portugal); MESTD (Serbia); MCIN/AEI and PCTI (Spain); MOSTR (Sri Lanka); Swiss Funding Agencies (Switzerland); MST (Taipei); MHESI and NSTDA (Thailand); TUBITAK and TENMAK (Turkey); NASU (Ukraine); STFC (United Kingdom); DOE and NSF (USA).
    
\hyphenation{Rachada-pisek} Individuals have received support from the Marie-Curie program and the European Research Council and Horizon 2020 Grant, contract Nos.\ 675440, 724704, 752730, 758316, 765710, 824093, 101115353, and COST Action CA16108 (European Union); the Leventis Foundation; the Alfred P.\ Sloan Foundation; the Alexander von Humboldt Foundation; the Science Committee, project no. 22rl-037 (Armenia); the Belgian Federal Science Policy Office; the Fonds pour la Formation \`a la Recherche dans l'Industrie et dans l'Agriculture (FRIA-Belgium); the Agentschap voor Innovatie door Wetenschap en Technologie (IWT-Belgium); the F.R.S.-FNRS and FWO (Belgium) under the ``Excellence of Science -- EOS" -- be.h project n.\ 30820817; the Beijing Municipal Science \& Technology Commission, No. Z191100007219010 and Fundamental Research Funds for the Central Universities (China); the Ministry of Education, Youth and Sports (MEYS) of the Czech Republic; the Shota Rustaveli National Science Foundation, grant FR-22-985 (Georgia); the Deutsche Forschungsgemeinschaft (DFG), under Germany's Excellence Strategy -- EXC 2121 ``Quantum Universe" -- 390833306, and under project number 400140256 - GRK2497; the Hellenic Foundation for Research and Innovation (HFRI), Project Number 2288 (Greece); the Hungarian Academy of Sciences, the New National Excellence Program - \'UNKP, the NKFIH research grants K 124845, K 124850, K 128713, K 128786, K 129058, K 131991, K 133046, K 138136, K 143460, K 143477, 2020-2.2.1-ED-2021-00181, and TKP2021-NKTA-64 (Hungary); the Council of Science and Industrial Research, India; ICSC -- National Research Center for High Performance Computing, Big Data and Quantum Computing, funded by the NextGenerationEU program (Italy); the Latvian Council of Science; the Ministry of Education and Science, project no. 2022/WK/14, and the National Science Center, contracts Opus 2021/41/B/ST2/01369 and 2021/43/B/ST2/01552 (Poland); the Funda\c{c}\~ao para a Ci\^encia e a Tecnologia, grant CEECIND/01334/2018 (Portugal); the National Priorities Research Program by Qatar National Research Fund; MCIN/AEI/10.13039/501100011033, ERDF ``a way of making Europe", and the Programa Estatal de Fomento de la Investigaci{\'o}n Cient{\'i}fica y T{\'e}cnica de Excelencia Mar\'{\i}a de Maeztu, grant MDM-2017-0765 and Programa Severo Ochoa del Principado de Asturias (Spain); the Chulalongkorn Academic into Its 2nd Century Project Advancement Project, and the National Science, Research and Innovation Fund via the Program Management Unit for Human Resources \& Institutional Development, Research and Innovation, grant B37G660013 (Thailand); the Kavli Foundation; the Nvidia Corporation; the SuperMicro Corporation; the Welch Foundation, contract C-1845; and the Weston Havens Foundation (USA).  
\end{acknowledgments}

\bibliography{auto_generated}

\providecommand{\href}[2]{#2}\begingroup\raggedright\begin{thebibliography}{10}%
\makeatletter
\providecommand{\hrefCMSnoop }[0]{\@secondoftwo}%
\makeatother
\providecommand{\doi}{\texttt{doi:}\begingroup \urlstyle{tt}\Url}

\bibitem{Super-Kamiokande:1998kpq}
\hrefCMSnoop {}{{Super-Kamiokande} Collaboration, ``Evidence for oscillation of
  atmospheric neutrinos'',} \textit{ Phys. Rev. Lett.} \textbf{ 81} (1998)
  1562,
  \href{http://dx.doi.org/10.1103/PhysRevLett.81.1562}{\doi{10.1103/PhysRevLett.81.1562}},
  \href{http://www.arXiv.org/abs/hep-ex/9807003}{\texttt{arXiv:hep-ex/9807003}}.

\bibitem{SNO:2002tuh}
\hrefCMSnoop {}{{SNO} Collaboration, ``Direct evidence for neutrino flavor
  transformation from neutral-current interactions in the {Sudbury Neutrino
  Observatory}'',} \textit{ Phys. Rev. Lett.} \textbf{ 89} (2002) 011301,
  \href{http://dx.doi.org/10.1103/PhysRevLett.89.011301}{\doi{10.1103/PhysRevLett.89.011301}},
  \href{http://www.arXiv.org/abs/nucl-ex/0204008}{\texttt{arXiv:nucl-ex/0204008}}.

\bibitem{KamLAND:2002uet}
\hrefCMSnoop {}{{KamLAND} Collaboration, ``First results from {KamLAND}:
  Evidence for reactor antineutrino disappearance'',} \textit{ Phys. Rev.
  Lett.} \textbf{ 90} (2003) 021802,
  \href{http://dx.doi.org/10.1103/PhysRevLett.90.021802}{\doi{10.1103/PhysRevLett.90.021802}},
  \href{http://www.arXiv.org/abs/hep-ex/0212021}{\texttt{arXiv:hep-ex/0212021}}.

\bibitem{Bilenky:2016pep}
\hrefCMSnoop {}{S.~Bilenky, ``Neutrino oscillations: From a historical
  perspective to the present status'',} \textit{ Nucl. Phys. B} \textbf{ 908}
  (2016) 2,
  \href{http://dx.doi.org/10.1016/j.nuclphysb.2016.01.025}{\doi{10.1016/j.nuclphysb.2016.01.025}},
  \href{http://www.arXiv.org/abs/1602.00170}{\texttt{arXiv:1602.00170}}.

\bibitem{RoyChoudhury:2019hls}
\hrefCMSnoop {}{S.~Roy~Choudhury and S.~Hannestad, ``Updated results on
  neutrino mass and mass hierarchy from cosmology with {Planck} 2018
  likelihoods'',} \textit{ JCAP} \textbf{ 07} (2020) 037,
  \href{http://dx.doi.org/10.1088/1475-7516/2020/07/037}{\doi{10.1088/1475-7516/2020/07/037}},
  \href{http://www.arXiv.org/abs/1907.12598}{\texttt{arXiv:1907.12598}}.

\bibitem{Ivanov:2019hqk}
\hrefCMSnoop {}{M.~Ivanov, M.~Simonovi{\'c}, and M.~Zaldarriaga, ``Cosmological
  parameters and neutrino masses from the final {Planck} and full-shape {BOSS}
  data'',} \textit{ Phys. Rev. D} \textbf{ 101} (2020) 083504,
  \href{http://dx.doi.org/10.1103/PhysRevD.101.083504}{\doi{10.1103/PhysRevD.101.083504}},
  \href{http://www.arXiv.org/abs/1912.08208}{\texttt{arXiv:1912.08208}}.

\bibitem{Formaggio:2021nfz}
\hrefCMSnoop {}{J.~Formaggio, A.~de~Gouv{\^e}a, and R.~Robertson, ``Direct
  measurements of neutrino mass'',} \textit{ Phys. Rept.} \textbf{ 914} (2021)
  1,
  \href{http://dx.doi.org/10.1016/j.physrep.2021.02.002}{\doi{10.1016/j.physrep.2021.02.002}},
  \href{http://www.arXiv.org/abs/2102.00594}{\texttt{arXiv:2102.00594}}.

\bibitem{Minkowski:1977sc}
\hrefCMSnoop {}{P.~Minkowski, ``${\PGm\to\Pe\PGg}$ at a rate of one out of
  $10^{9}$ muon decays?'',} \textit{ Phys. Lett. B} \textbf{ 67} (1977) 421,
  \href{http://dx.doi.org/10.1016/0370-2693(77)90435-X}{\doi{10.1016/0370-2693(77)90435-X}}.

\bibitem{Yanagida:1979as}
\hrefCMSnoop {}{T.~Yanagida, ``Horizontal symmetry and masses of neutrinos'',}
  \textit{ Prog. of Theo. Phys.} \textbf{ 64} (1980) 1103.

\bibitem{Gell-Mann:1979vob}
\hrefCMSnoop {}{M.~Gell-Mann, P.~Ramond, and R.~Slansky, ``Complex spinors and
  unified theories'',} in \textit{ Supergravity}, p.~315.
\newblock North Holland Publishing, 1979.
\newblock \href{http://www.arXiv.org/abs/1306.4669}{\texttt{arXiv:1306.4669}}.

\bibitem{Glashow:1979nm}
\hrefCMSnoop {}{S.~Glashow, ``The future of elementary particle physics'',}
  \textit{ NATO Sci. Ser. B} \textbf{ 61} (1980) 687,
  \href{http://dx.doi.org/10.1007/978-1-4684-7197-7_15}{\doi{10.1007/978-1-4684-7197-7_15}}.

\bibitem{Mohapatra:1979ia}
\hrefCMSnoop {}{R.~N. Mohapatra and G.~Senjanovi{\'c}, ``Neutrino mass and
  spontaneous parity nonconservation'',} \textit{ Phys. Rev. Lett.} \textbf{
  44} (1980) 912,
  \href{http://dx.doi.org/10.1103/PhysRevLett.44.912}{\doi{10.1103/PhysRevLett.44.912}}.

\bibitem{Schechter:1980gr}
\hrefCMSnoop {}{J.~Schechter and J.~W.~F. Valle, ``Neutrino masses in
  $\mathrm{SU}(2)\bigotimes\mathrm{U}(1)$ theories'',} \textit{ Phys. Rev. D}
  \textbf{ 22} (1980) 2227,
  \href{http://dx.doi.org/10.1103/PhysRevD.22.2227}{\doi{10.1103/PhysRevD.22.2227}}.

\bibitem{Shrock:1980ct}
\hrefCMSnoop {}{R.~E. Shrock, ``General theory of weak leptonic and
  semileptonic decays. {I}. {L}eptonic pseudoscalar meson decays, with
  associated tests for, and bounds on, neutrino masses and lepton mixing'',}
  \textit{ Phys. Rev. D} \textbf{ 24} (1981) 1232,
  \href{http://dx.doi.org/10.1103/PhysRevD.24.1232}{\doi{10.1103/PhysRevD.24.1232}}.

\bibitem{Cai:2017mow}
\hrefCMSnoop {}{Y.~Cai, T.~Han, T.~Li, and R.~Ruiz, ``Lepton number violation:
  Seesaw models and their collider tests'',} \textit{ Front. Phys.} \textbf{ 6}
  (2018) 40,
  \href{http://dx.doi.org/10.3389/fphy.2018.00040}{\doi{10.3389/fphy.2018.00040}},
  \href{http://www.arXiv.org/abs/1711.02180}{\texttt{arXiv:1711.02180}}.

\bibitem{Maki:1962mu}
\hrefCMSnoop {}{Z.~Maki, M.~Nakagawa, and S.~Sakata, ``Remarks on the unified
  model of elementary particles'',} \textit{ Prog. Theor. Phys.} \textbf{ 28}
  (1962) 870,
  \href{http://dx.doi.org/10.1143/PTP.28.870}{\doi{10.1143/PTP.28.870}}.

\bibitem{Pontecorvo:1967fh}
\hrefCMSnoop {}{B.~Pontecorvo, ``Neutrino experiments and the problem of
  conservation of leptonic charge'',} \textit{ Zh. Eksp. Teor. Fiz.} \textbf{
  53} (1967) 1717.

\bibitem{Das_2016}
\hrefCMSnoop {}{A.~Das, P.~Konar, and S.~Majhi, ``{Production of Heavy neutrino
  in next-to-leading order QCD at the LHC and beyond}'',} \textit{ JHEP}
  \textbf{ 06} (2016) 019,
  \href{http://dx.doi.org/10.1007/JHEP06(2016)019}{\doi{10.1007/JHEP06(2016)019}},
  \href{http://www.arXiv.org/abs/1604.00608}{\texttt{arXiv:1604.00608}}.

\bibitem{Das_2017}
\hrefCMSnoop {}{A.~Das and N.~Okada, ``{Bounds on heavy Majorana neutrinos in
  type-I seesaw and implications for collider searches}'',} \textit{ Phys.
  Lett. B} \textbf{ 774} (2017) 32,
  \href{http://dx.doi.org/10.1016/j.physletb.2017.09.042}{\doi{10.1016/j.physletb.2017.09.042}},
  \href{http://www.arXiv.org/abs/1702.04668}{\texttt{arXiv:1702.04668}}.

\bibitem{cottin2023long}
G.~Cottin\hrefCMSnoop {}{ { et~al.}, ``{Long-lived heavy neutral leptons with a
  displaced shower signature at CMS}'',} \textit{ JHEP} \textbf{ 02} (2023)
  011,
  \href{http://dx.doi.org/10.1007/JHEP02(2023)011}{\doi{10.1007/JHEP02(2023)011}},
  \href{http://www.arXiv.org/abs/2210.17446}{\texttt{arXiv:2210.17446}}.

\bibitem{Bondarenko:2018ptm}
\hrefCMSnoop {}{K.~Bondarenko, A.~Boyarsky, D.~Gorbunov, and O.~Ruchayskiy,
  ``Phenomenology of {\GeVns}-scale heavy neutral leptons'',} \textit{ JHEP}
  \textbf{ 11} (2018) 032,
  \href{http://dx.doi.org/10.1007/JHEP11(2018)032}{\doi{10.1007/JHEP11(2018)032}},
  \href{http://www.arXiv.org/abs/1805.08567}{\texttt{arXiv:1805.08567}}.

\bibitem{Fukugita:1986hr}
\hrefCMSnoop {}{M.~Fukugita and T.~Yanagida, ``Baryogenesis without grand
  unification'',} \textit{ Phys. Lett. B} \textbf{ 174} (1986) 45,
  \href{http://dx.doi.org/10.1016/0370-2693(86)91126-3}{\doi{10.1016/0370-2693(86)91126-3}}.

\bibitem{Chun:2017spz}
E.~Chun\hrefCMSnoop {}{ { et~al.}, ``Probing leptogenesis'',} \textit{ Int. J.
  Mod. Phys. A} \textbf{ 33} (2018) 1842005,
  \href{http://dx.doi.org/10.1142/S0217751X18420058}{\doi{10.1142/S0217751X18420058}},
  \href{http://www.arXiv.org/abs/1711.02865}{\texttt{arXiv:1711.02865}}.

\bibitem{boyarsky2019sterile}
A.~Boyarsky\hrefCMSnoop {}{ { et~al.}, ``Sterile neutrino dark matter'',}
  \textit{ Prog. Part. Nucl. Phys.} \textbf{ 104} (2019) 1,
  \href{http://dx.doi.org/10.1016/j.ppnp.2018.07.004}{\doi{10.1016/j.ppnp.2018.07.004}},
  \href{http://www.arXiv.org/abs/1807.07938v2}{\texttt{arXiv:1807.07938v2}}.

\bibitem{Cirigliano_2021}
V.~Cirigliano\hrefCMSnoop {}{ { et~al.}, ``{Leptonic anomalous magnetic moments
  in \ensuremath{\nu} SMEFT}'',} \textit{ JHEP} \textbf{ 08} (2021) 103,
  \href{http://dx.doi.org/10.1007/JHEP08(2021)103}{\doi{10.1007/JHEP08(2021)103}},
  \href{http://www.arXiv.org/abs/2105.11462}{\texttt{arXiv:2105.11462}}.

\bibitem{Muong-2:2023}
\hrefCMSnoop {}{{Muon g-2} Collaboration, ``{Measurement of the Positive Muon
  Anomalous Magnetic Moment to 0.20~ppm}'',} \textit{ Phys. Rev. Lett.}
  \textbf{ 131} (2023) 161802,
  \href{http://dx.doi.org/10.1103/PhysRevLett.131.161802}{\doi{10.1103/PhysRevLett.131.161802}},
  \href{http://www.arXiv.org/abs/2308.06230}{\texttt{arXiv:2308.06230}}.

\bibitem{BEBC}
\hrefCMSnoop {}{{WA66} Collaboration, ``Search for heavy neutrino decays in the
  {BEBC} beam dump experiment'',} \textit{ Phys. Lett. B} \textbf{ 160} (1985)
  207,
  \href{http://dx.doi.org/10.1016/0370-2693(85)91493-5}{\doi{10.1016/0370-2693(85)91493-5}}.

\bibitem{CHARM:1985nku}
\hrefCMSnoop {}{{CHARM} Collaboration, ``A search for decays of heavy neutrinos
  in the mass range {0.5--2.8\GeV}'',} \textit{ Phys. Lett. B} \textbf{ 166}
  (1986) 473,
  \href{http://dx.doi.org/10.1016/0370-2693(86)91601-1}{\doi{10.1016/0370-2693(86)91601-1}}.

\bibitem{delphi1997search}
\hrefCMSnoop {}{{{DELPHI}} Collaboration, ``Search for neutral heavy leptons
  produced in {Z} decays'',} \textit{ Z. Phys. C} \textbf{ 74} (1997) 57,
  \href{http://dx.doi.org/10.1007/s002880050370}{\doi{10.1007/s002880050370}}.
  [Erratum: \DOI{10.1007/BF03546181}].

\bibitem{NuTeV}
\hrefCMSnoop {}{{NuTeV(E815)} Collaboration, ``Search for neutral heavy leptons
  in a high-energy neutrino beam'',} \textit{ Phys. Rev. Lett.} \textbf{ 83}
  (1999) 4943,
  \href{http://dx.doi.org/10.1103/PhysRevLett.83.4943}{\doi{10.1103/PhysRevLett.83.4943}},
  \href{http://www.arXiv.org/abs/hep-ex/9908011}{\texttt{arXiv:hep-ex/9908011}}.

\bibitem{Belle}
\hrefCMSnoop {}{{Belle} Collaboration, ``{Search for heavy neutrinos at
  Belle}'',} \textit{ Phys. Rev. D} \textbf{ 87} (2013) 071102,
  \href{http://dx.doi.org/10.1103/PhysRevD.87.071102}{\doi{10.1103/PhysRevD.87.071102}},
  \href{http://www.arXiv.org/abs/1301.1105}{\texttt{arXiv:1301.1105}}.
  [Erratum: \DOI{10.1103/PhysRevD.95.099903}].

\bibitem{Deppisch:2015qwa}
\hrefCMSnoop {}{F.~F. Deppisch, P.~S. Bhupal~Dev, and A.~Pilaftsis, ``Neutrinos
  and collider physics'',} \textit{ New J. Phys.} \textbf{ 17} (2015) 075019,
  \href{http://dx.doi.org/10.1088/1367-2630/17/7/075019}{\doi{10.1088/1367-2630/17/7/075019}},
  \href{http://www.arXiv.org/abs/1502.06541}{\texttt{arXiv:1502.06541}}.

\bibitem{Beacham:2019nyx}
\hrefCMSnoop {}{J.~Beacham { et~al.}, ``Physics beyond colliders at {CERN}:
  Beyond the standard model working group report'',} \textit{ J. Phys. G}
  \textbf{ 47} (2020) 010501,
  \href{http://dx.doi.org/10.1088/1361-6471/ab4cd2}{\doi{10.1088/1361-6471/ab4cd2}},
  \href{http://www.arXiv.org/abs/1901.09966}{\texttt{arXiv:1901.09966}}.

\bibitem{ATLAS:2019kpx}
\hrefCMSnoop {}{{ATLAS Collaboration}, ``{Search for heavy neutral leptons in
  decays of W~bosons produced in 13 TeV pp collisions using prompt and
  displaced signatures with the ATLAS detector}'',} \textit{ JHEP} \textbf{ 10}
  (2019) 265,
  \href{http://dx.doi.org/10.1007/JHEP10(2019)265}{\doi{10.1007/JHEP10(2019)265}},
  \href{http://www.arXiv.org/abs/1905.09787}{\texttt{arXiv:1905.09787}}.

\bibitem{ATLAS:2022atq}
\hrefCMSnoop {}{{ATLAS Collaboration}, ``{Search for heavy neutral leptons in
  decays of $W$ bosons using a dilepton displaced vertex in $\sqrt{s}=13$ TeV
  pp collisions with the ATLAS detector}'',} \textit{ Phys. Rev. Lett.}
  \textbf{ 131} (2023) 061803,
  \href{http://dx.doi.org/10.1103/PhysRevLett.131.061803}{\doi{10.1103/PhysRevLett.131.061803}},
  \href{http://www.arXiv.org/abs/2204.11988}{\texttt{arXiv:2204.11988}}.

\bibitem{CMS:2018iaf}
\hrefCMSnoop {}{{CMS Collaboration}, ``{Search for heavy neutral leptons in
  events with three charged leptons in proton-proton collisions at $\sqrt{s} =
  13$ TeV}'',} \textit{ Phys. Rev. Lett.} \textbf{ 120} (2018) 221801,
  \href{http://dx.doi.org/10.1103/PhysRevLett.120.221801}{\doi{10.1103/PhysRevLett.120.221801}},
  \href{http://www.arXiv.org/abs/1802.02965}{\texttt{arXiv:1802.02965}}.

\bibitem{CMS:2018jxx}
\hrefCMSnoop {}{{CMS Collaboration}, ``{Search for heavy Majorana neutrinos in
  same-sign dilepton channels in proton-proton collisions at $ \sqrt{s}=13$
  TeV}'',} \textit{ JHEP} \textbf{ 01} (2019) 122,
  \href{http://dx.doi.org/10.1007/JHEP01(2019)122}{\doi{10.1007/JHEP01(2019)122}},
  \href{http://www.arXiv.org/abs/1806.10905}{\texttt{arXiv:1806.10905}}.

\bibitem{CMS:2022fut}
\hrefCMSnoop {}{{CMS Collaboration}, ``Search for long-lived heavy neutral
  leptons with displaced vertices in proton-proton collisions at
  {$\sqrt{s}=13\TeV$}'',} \textit{ JHEP} \textbf{ 07} (2022) 081,
  \href{http://dx.doi.org/10.1007/JHEP07(2022)081}{\doi{10.1007/JHEP07(2022)081}},
  \href{http://www.arXiv.org/abs/2201.05578}{\texttt{arXiv:2201.05578}}.

\bibitem{LHCb:2020wxx}
\hrefCMSnoop {}{{LHCb Collaboration}, ``{Search for heavy neutral leptons in
  $\textrm{W}^+\to\mu^{+}\mu^{\pm}\textrm{jet}$ decays}'',} \textit{ Eur. Phys.
  J. C} \textbf{ 81} (2021) 248,
  \href{http://dx.doi.org/10.1140/epjc/s10052-021-08973-5}{\doi{10.1140/epjc/s10052-021-08973-5}},
  \href{http://www.arXiv.org/abs/2011.05263}{\texttt{arXiv:2011.05263}}.

\bibitem{LHCb:2020akw}
\hrefCMSnoop {}{{LHCb Collaboration}, ``{Search for long-lived particles
  decaying to $\textrm{e}^\pm \mu^\mp \nu$}'',} \textit{ Eur. Phys. J. C}
  \textbf{ 81} (2021) 261,
  \href{http://dx.doi.org/10.1140/epjc/s10052-021-08994-0}{\doi{10.1140/epjc/s10052-021-08994-0}},
  \href{http://www.arXiv.org/abs/2012.02696}{\texttt{arXiv:2012.02696}}.

\bibitem{CMS:2023jqi}
\hrefCMSnoop {}{{CMS Collaboration}, ``{Search for long-lived heavy neutral
  leptons with lepton flavour conserving or violating decays to a jet and a
  charged lepton}'',} 12, 2023.
  \href{http://www.arXiv.org/abs/2312.07484}{\texttt{arXiv:2312.07484}}.
  Submitted to \emph{JHEP}.

\bibitem{CMS:2021juv}
\hrefCMSnoop {}{{CMS Collaboration}, ``Search for long-lived particles decaying
  in the {CMS} endcap muon detectors in proton-proton collisions at
  {$\sqrt{s}=13\TeV$}'',} \textit{ Phys. Rev. Lett.} \textbf{ 127} (2021)
  261804,
  \href{http://dx.doi.org/10.1103/PhysRevLett.127.261804}{\doi{10.1103/PhysRevLett.127.261804}},
  \href{http://www.arXiv.org/abs/2107.04838}{\texttt{arXiv:2107.04838}}.

\bibitem{hepdata}
\hrefCMSnoop {}{}{HEPD}ata record for this analysis, 2023.
\newblock
  \href{http://dx.doi.org/10.17182/hepdata.146757}{\doi{10.17182/hepdata.146757}}.

\bibitem{CMS:2009lnj}
\hrefCMSnoop {}{{CMS Collaboration}, ``{Performance of the CMS Drift Tube
  Chambers with Cosmic Rays}'',} \textit{ JINST} \textbf{ 5} (2010) T03015,
  \href{http://dx.doi.org/10.1088/1748-0221/5/03/T03015}{\doi{10.1088/1748-0221/5/03/T03015}},
  \href{http://www.arXiv.org/abs/0911.4855}{\texttt{arXiv:0911.4855}}.

\bibitem{CMS:2020cmk}
\hrefCMSnoop {}{{CMS Collaboration}, ``{Performance of the CMS Level-1 trigger
  in proton-proton collisions at $\sqrt{s} = 13\TeV$}'',} \textit{ JINST}
  \textbf{ 15} (2020) P10017,
  \href{http://dx.doi.org/10.1088/1748-0221/15/10/P10017}{\doi{10.1088/1748-0221/15/10/P10017}},
  \href{http://www.arXiv.org/abs/2006.10165}{\texttt{arXiv:2006.10165}}.

\bibitem{CMS:2016ngn}
\hrefCMSnoop {}{{CMS Collaboration}, ``{The CMS trigger system}'',} \textit{
  JINST} \textbf{ 12} (2017) P01020,
  \href{http://dx.doi.org/10.1088/1748-0221/12/01/P01020}{\doi{10.1088/1748-0221/12/01/P01020}},
  \href{http://www.arXiv.org/abs/1609.02366}{\texttt{arXiv:1609.02366}}.

\bibitem{CMS:2008xjf}
\hrefCMSnoop {}{{CMS Collaboration}, ``The {CMS} experiment at the {CERN}
  {LHC}'',} \textit{ JINST} \textbf{ 3} (2008) S08004,
  \href{http://dx.doi.org/10.1088/1748-0221/3/08/S08004}{\doi{10.1088/1748-0221/3/08/S08004}}.

\bibitem{POWHEG1}
\hrefCMSnoop {}{P.~Nason, ``A new method for combining {NLO QCD} with shower
  {Monte Carlo} algorithms'',} \textit{ JHEP} \textbf{ 11} (2004) 040,
  \href{http://dx.doi.org/10.1088/1126-6708/2004/11/040}{\doi{10.1088/1126-6708/2004/11/040}},
\href{http://www.arXiv.org/abs/hep-ph/0409146}{\texttt{arXiv:hep-ph/0409146}}.

\bibitem{POWHEG2}
\hrefCMSnoop {}{S.~Frixione, P.~Nason, and C.~Oleari, ``Matching {NLO QCD}
  computations with parton shower simulations: the {POWHEG} method'',} \textit{
  JHEP} \textbf{ 11} (2007) 070,
  \href{http://dx.doi.org/10.1088/1126-6708/2007/11/070}{\doi{10.1088/1126-6708/2007/11/070}},
\href{http://www.arXiv.org/abs/0709.2092}{\texttt{arXiv:0709.2092}}.

\bibitem{POWHEG3}
\hrefCMSnoop {}{S.~Alioli, P.~Nason, C.~Oleari, and E.~Re, ``A general
  framework for implementing {NLO} calculations in shower {Monte Carlo}
  programs: the {\POWHEG box}'',} \textit{ JHEP} \textbf{ 06} (2010) 043,
  \href{http://dx.doi.org/10.1007/JHEP06(2010)043}{\doi{10.1007/JHEP06(2010)043}},
\href{http://www.arXiv.org/abs/1002.2581}{\texttt{arXiv:1002.2581}}.

\bibitem{Gavin:2010az}
\hrefCMSnoop {}{R.~Gavin, Y.~Li, F.~Petriello, and S.~Quackenbush,
  ``{\textsc{FEWZ}} 2.0: {A} code for hadronic {\PZ} production at
  next-to-next-to-leading order'',} \textit{ Comput. Phys. Commun.} \textbf{
  182} (2011) 2388,
  \href{http://dx.doi.org/10.1016/j.cpc.2011.06.008}{\doi{10.1016/j.cpc.2011.06.008}},
\href{http://www.arXiv.org/abs/1011.3540}{\texttt{arXiv:1011.3540}}.

\bibitem{Gavin:2012sy}
\hrefCMSnoop {}{R.~Gavin, Y.~Li, F.~Petriello, and S.~Quackenbush, ``{\PW}
  physics at the {LHC} with {\textsc{fewz} 2.1}'',} \textit{ Comput. Phys.
  Commun.} \textbf{ 184} (2013) 208,
  \href{http://dx.doi.org/10.1016/j.cpc.2012.09.005}{\doi{10.1016/j.cpc.2012.09.005}},
\href{http://www.arXiv.org/abs/1201.5896}{\texttt{arXiv:1201.5896}}.

\bibitem{Alwall_2014}
J.~Alwall\hrefCMSnoop {}{ { et~al.}, ``The automated computation of tree-level
  and next-to-leading order differential cross sections, and their matching to
  parton shower simulations'',} \textit{ JHEP} \textbf{ 07} (2014) 079,
  \href{http://dx.doi.org/10.1007/JHEP07(2014)079}{\doi{10.1007/JHEP07(2014)079}},
  \href{http://www.arXiv.org/abs/1405.0301}{\texttt{arXiv:1405.0301}}.

\bibitem{Alva_2015}
\hrefCMSnoop {}{D.~Alva, T.~Han, and R.~Ruiz, ``{Heavy Majorana neutrinos from
  $\PW\gamma$ fusion at hadron colliders}'',} \textit{ JHEP} \textbf{ 02}
  (2015) 072,
  \href{http://dx.doi.org/10.1007/JHEP02(2015)072}{\doi{10.1007/JHEP02(2015)072}},
  \href{http://www.arXiv.org/abs/1411.7305}{\texttt{arXiv:1411.7305}}.

\bibitem{Degrande_2016}
\hrefCMSnoop {}{C.~Degrande, O.~Mattelaer, R.~Ruiz, and J.~Turner,
  ``Fully-automated precision predictions for heavy neutrino production
  mechanisms at hadron colliders'',} \textit{ Phys. Rev. D} \textbf{ 94} (2016)
  053002,
  \href{http://dx.doi.org/10.1103/PhysRevD.94.053002}{\doi{10.1103/PhysRevD.94.053002}},
  \href{http://www.arXiv.org/abs/1602.06957}{\texttt{arXiv:1602.06957}}.

\bibitem{Pascoli_2019}
\hrefCMSnoop {}{S.~Pascoli, R.~Ruiz, and C.~Weiland, ``{Heavy neutrinos with
  dynamic jet vetoes: Multilepton searches at $\sqrt{s}=14$, 27, and
  100\TeV}'',} \textit{ JHEP} \textbf{ 06} (2019) 049,
  \href{http://dx.doi.org/10.1007/JHEP06(2019)049}{\doi{10.1007/JHEP06(2019)049}},
  \href{http://www.arXiv.org/abs/1812.08750}{\texttt{arXiv:1812.08750}}.

\bibitem{Sj_strand_2015}
T.~Sj{\"o}strand\hrefCMSnoop {}{ { et~al.}, ``An introduction to
  {\PYTHIA8.2}'',} \textit{ Comp. Phys. Comm.} \textbf{ 191} (2015) 159,
  \href{http://dx.doi.org/10.1016/j.cpc.2015.01.024}{\doi{10.1016/j.cpc.2015.01.024}},
  \href{http://www.arXiv.org/abs/1410.3012}{\texttt{arXiv:1410.3012}}.

\bibitem{FEWZ}
\hrefCMSnoop {}{Y.~Li and F.~Petriello, ``{Combining QCD and electroweak
  corrections to dilepton production in \textsc{FEWZ}}'',} \textit{ Phys. Rev.
  D} \textbf{ 86} (2012) 094034,
  \href{http://dx.doi.org/10.1103/PhysRevD.86.094034}{\doi{10.1103/PhysRevD.86.094034}},
  \href{http://www.arXiv.org/abs/1208.5967}{\texttt{arXiv:1208.5967}}.

\bibitem{DYTURBO}
S.~Camarda\hrefCMSnoop {}{ { et~al.}, ``{\textsc{DYTurbo}: Fast predictions for
  Drell-Yan processes}'',} \textit{ Eur. Phys. J. C} \textbf{ 80} (2020) 251,
  \href{http://dx.doi.org/10.1140/epjc/s10052-020-7757-5}{\doi{10.1140/epjc/s10052-020-7757-5}},
  \href{http://www.arXiv.org/abs/1910.07049}{\texttt{arXiv:1910.07049}}.

\bibitem{Camarda_2021}
\hrefCMSnoop {}{S.~Camarda, L.~Cieri, and G.~Ferrera, ``Drell\textendash{}yan
  lepton-pair production: ${q}_{T}$ resummation at $\mathrm{N}^{3}\mathrm{LL}$
  accuracy and fiducial cross sections at $\mathrm{N}^{3}\mathrm{LO}$'',}
  \textit{ Phys. Rev. D} \textbf{ 104} (2021) L111503,
  \href{http://dx.doi.org/10.1103/PhysRevD.104.L111503}{\doi{10.1103/PhysRevD.104.L111503}},
  \href{http://www.arXiv.org/abs/2103.04974v3}{\texttt{arXiv:2103.04974v3}}.

\bibitem{AGOSTINELLI2003250}
\hrefCMSnoop {}{{GEANT4} Collaboration, ``{\GEANTfour}---a simulation
  toolkit'',} \textit{ Nucl. Instrum. Meth. A} \textbf{ 506} (2003) 250,
  \href{http://dx.doi.org/10.1016/S0168-9002(03)01368-8}{\doi{10.1016/S0168-9002(03)01368-8}}.

\bibitem{CMS-LUM-17-003}
\hrefCMSnoop {}{{CMS Collaboration}, ``Precision luminosity measurement in
  proton-proton collisions at {$\sqrt{s}=13\TeV$} in 2015 and 2016 at {CMS}'',}
  \textit{ Eur. Phys. J. C} \textbf{ 81} (2021) 800,
  \href{http://dx.doi.org/10.1140/epjc/s10052-021-09538-2}{\doi{10.1140/epjc/s10052-021-09538-2}},
  \href{http://www.arXiv.org/abs/2104.01927}{\texttt{arXiv:2104.01927}}.

\bibitem{CMS-PAS-LUM-17-004}
\href {https://cds.cern.ch/record/2621960/}{{CMS Collaboration}, ``{CMS}
  luminosity measurement for the 2017 data-taking period at {$\sqrt{s} =
  13\TeV$}'',} CMS Physics Analysis Summary CMS-PAS-LUM-17-004, 2018.

\bibitem{CMS-PAS-LUM-18-002}
\href {https://cds.cern.ch/record/2676164/}{{CMS Collaboration}, ``{CMS}
  luminosity measurement for the 2018 data-taking period at {$\sqrt{s} =
  13\TeV$}'',} CMS Physics Analysis Summary CMS-PAS-LUM-18-002, 2019.

\bibitem{Khachatryan:2015pea}
\hrefCMSnoop {}{{CMS Collaboration}, ``Event generator tunes obtained from
  underlying event and multiparton scattering measurements'',} \textit{ Eur.
  Phys. J. C} \textbf{ 76} (2016) 155,
  \href{http://dx.doi.org/10.1140/epjc/s10052-016-3988-x}{\doi{10.1140/epjc/s10052-016-3988-x}},
  \href{http://www.arXiv.org/abs/1512.00815}{\texttt{arXiv:1512.00815}}.

\bibitem{CP5}
\hrefCMSnoop {}{{CMS Collaboration}, ``Extraction and validation of a new set
  of {CMS \PYTHIA8} tunes from underlying-event measurements'',} \textit{ Eur.
  Phys. J. C} \textbf{ 80} (2020) 4,
  \href{http://dx.doi.org/10.1140/epjc/s10052-019-7499-4}{\doi{10.1140/epjc/s10052-019-7499-4}},
  \href{http://www.arXiv.org/abs/1903.12179}{\texttt{arXiv:1903.12179}}.

\bibitem{Ball:2014uwa}
\hrefCMSnoop {}{{NNPDF} Collaboration, ``Parton distributions for the {LHC Run
  II}'',} \textit{ JHEP} \textbf{ 04} (2015) 040,
  \href{http://dx.doi.org/10.1007/JHEP04(2015)040}{\doi{10.1007/JHEP04(2015)040}},
  \href{http://www.arXiv.org/abs/1410.8849}{\texttt{arXiv:1410.8849}}.

\bibitem{Ball:2017nwa}
\hrefCMSnoop {}{{NNPDF} Collaboration, ``Parton distributions from
  high-precision collider data'',} \textit{ Eur. Phys. J. C} \textbf{ 77}
  (2017) 663,
  \href{http://dx.doi.org/10.1140/epjc/s10052-017-5199-5}{\doi{10.1140/epjc/s10052-017-5199-5}},
  \href{http://www.arXiv.org/abs/1706.00428}{\texttt{arXiv:1706.00428}}.

\bibitem{Sirunyan:2017ulk}
\hrefCMSnoop {}{{CMS Collaboration}, ``Particle-flow reconstruction and global
  event description with the {CMS} detector'',} \textit{ JINST} \textbf{ 12}
  (2017) P10003,
  \href{http://dx.doi.org/10.1088/1748-0221/12/10/P10003}{\doi{10.1088/1748-0221/12/10/P10003}},
  \href{http://www.arXiv.org/abs/1706.04965}{\texttt{arXiv:1706.04965}}.

\bibitem{CMS:2020uim}
\hrefCMSnoop {}{{CMS Collaboration}, ``Electron and photon reconstruction and
  identification with the {CMS} experiment at the {CERN} {LHC}'',} \textit{
  JINST} \textbf{ 16} (2021) P05014,
  \href{http://dx.doi.org/10.1088/1748-0221/16/05/P05014}{\doi{10.1088/1748-0221/16/05/P05014}},
  \href{http://www.arXiv.org/abs/2012.06888}{\texttt{arXiv:2012.06888}}.

\bibitem{CMS:2018rym}
\hrefCMSnoop {}{{CMS Collaboration}, ``{Performance of the CMS muon detector
  and muon reconstruction with proton-proton collisions at $\sqrt{s}=
  13\TeV$}'',} \textit{ JINST} \textbf{ 13} (2018) P06015,
  \href{http://dx.doi.org/10.1088/1748-0221/13/06/P06015}{\doi{10.1088/1748-0221/13/06/P06015}},
  \href{http://www.arXiv.org/abs/1804.04528}{\texttt{arXiv:1804.04528}}.

\bibitem{CMS:2015myp}
\hrefCMSnoop {}{{CMS Collaboration}, ``Performance of photon reconstruction and
  identification with the {CMS} detector in proton-proton collisions at
  {$\sqrt{s} = 8\TeV$}'',} \textit{ JINST} \textbf{ 10} (2015) P08010,
  \href{http://dx.doi.org/10.1088/1748-0221/10/08/P08010}{\doi{10.1088/1748-0221/10/08/P08010}},
  \href{http://www.arXiv.org/abs/1502.02702}{\texttt{arXiv:1502.02702}}.

\bibitem{Cacciari:2008gp}
\hrefCMSnoop {}{M.~Cacciari, G.~P. Salam, and G.~Soyez, ``The anti-{\kt} jet
  clustering algorithm'',} \textit{ JHEP} \textbf{ 04} (2008) 063,
  \href{http://dx.doi.org/10.1088/1126-6708/2008/04/063}{\doi{10.1088/1126-6708/2008/04/063}},
  \href{http://www.arXiv.org/abs/0802.1189}{\texttt{arXiv:0802.1189}}.

\bibitem{Cacciari:2011ma}
\hrefCMSnoop {}{M.~Cacciari, G.~P. Salam, and G.~Soyez, ``{\FASTJET} user
  manual'',} \textit{ Eur. Phys. J. C} \textbf{ 72} (2012) 1896,
  \href{http://dx.doi.org/10.1140/epjc/s10052-012-1896-2}{\doi{10.1140/epjc/s10052-012-1896-2}},
  \href{http://www.arXiv.org/abs/1111.6097}{\texttt{arXiv:1111.6097}}.

\bibitem{CMS:2020ebo}
\hrefCMSnoop {}{{CMS Collaboration}, ``{Pileup mitigation at CMS in 13\TeV
  data}'',} \textit{ JINST} \textbf{ 15} (2020) P09018,
  \href{http://dx.doi.org/10.1088/1748-0221/15/09/P09018}{\doi{10.1088/1748-0221/15/09/P09018}},
  \href{http://www.arXiv.org/abs/2003.00503}{\texttt{arXiv:2003.00503}}.

\bibitem{Khachatryan:2016kdb}
\hrefCMSnoop {}{{CMS Collaboration}, ``{Jet energy scale and resolution in the
  CMS experiment in pp collisions at 8 TeV}'',} \textit{ JINST} \textbf{ 12}
  (2017) P02014,
  \href{http://dx.doi.org/10.1088/1748-0221/12/02/P02014}{\doi{10.1088/1748-0221/12/02/P02014}},
  \href{http://www.arXiv.org/abs/1607.03663}{\texttt{arXiv:1607.03663}}.

\bibitem{CMS:2019ctu}
\hrefCMSnoop {}{{CMS Collaboration}, ``Performance of missing transverse
  momentum reconstruction in proton-proton collisions at $\sqrt{s} = 13$\,{TeV}
  using the {CMS} detector'',} \textit{ JINST} \textbf{ 14} (2019) P07004,
  \href{http://dx.doi.org/10.1088/1748-0221/14/07/P07004}{\doi{10.1088/1748-0221/14/07/P07004}},
  \href{http://www.arXiv.org/abs/1903.06078}{\texttt{arXiv:1903.06078}}.

\bibitem{CMS-TDR-15-02}
\href {http://cds.cern.ch/record/2020886}{{CMS Collaboration}, ``Technical
  proposal for the {Phase-II} upgrade of the {Compact Muon Solenoid}'',} CMS
  Technical Proposal CERN-LHCC-2015-010, CMS-TDR-15-02, 2015.

\bibitem{dbscan}
\href {https://www.aaai.org/Papers/KDD/1996/KDD96-037.pdf}{M.~Ester, H.-P.
  Kriegel, J.~Sander, and X.~Xu, ``{A density-based algorithm for discovering
  clusters in large spatial databases with noise}'',} in \textit{ Proc. 2nd
  Int. Conf. on Knowledge Discovery and Data Mining}, p.~226.
\newblock Association for the Advancement of Artificial Intelligence, 1996.

\bibitem{btag}
\hrefCMSnoop {}{{CMS Collaboration}, ``{Identification of b-quark jets with the
  CMS experiment}'',} \textit{ JINST} \textbf{ 8} (2013) P04013,
  \href{http://dx.doi.org/10.1088/1748-0221/8/04/P04013}{\doi{10.1088/1748-0221/8/04/P04013}},
  \href{http://www.arXiv.org/abs/1211.4462}{\texttt{arXiv:1211.4462}}.

\bibitem{BTV-16-002}
\hrefCMSnoop {}{{CMS Collaboration}, ``{Identification of heavy-flavour jets
  with the CMS detector in pp collisions at 13 TeV}'',} \textit{ JINST}
  \textbf{ 13} (2018) P05011,
  \href{http://dx.doi.org/10.1088/1748-0221/13/05/P05011}{\doi{10.1088/1748-0221/13/05/P05011}},
  \href{http://www.arXiv.org/abs/1712.07158}{\texttt{arXiv:1712.07158}}.

\bibitem{JUNK1999435}
\hrefCMSnoop {}{T.~Junk, ``{Confidence level computation for combining searches
  with small statistics}'',} \textit{ Nucl. Instrum. Meth. A} \textbf{ 434}
  (1999) 435,
  \href{http://dx.doi.org/10.1016/S0168-9002(99)00498-2}{\doi{10.1016/S0168-9002(99)00498-2}},
  \href{http://www.arXiv.org/abs/hep-ex/9902006}{\texttt{arXiv:hep-ex/9902006}}.

\bibitem{Read_2002}
\hrefCMSnoop {}{A.~L. Read, ``{Presentation of search results: the \CLs
  technique}'',} \textit{ J. Phys. G} \textbf{ 28} (2002) 2693,
  \href{http://dx.doi.org/10.1088/0954-3899/28/10/313}{\doi{10.1088/0954-3899/28/10/313}}.

\bibitem{CMS-NOTE-2011-005}
\href {https://cds.cern.ch/record/1379837}{{ATLAS and CMS Collaborations, and
  the LHC Higgs Combination Group}, ``Procedure for the {LHC} {Higgs} boson
  search combination in {Summer} 2011'',} Technical Report CMS-NOTE-2011-005,
  ATL-PHYS-PUB-2011-11, 2011.

\bibitem{Cowan:2010js}
\hrefCMSnoop {}{G.~Cowan, K.~Cranmer, E.~Gross, and O.~Vitells, ``Asymptotic
  formulae for likelihood-based tests of new physics'',} \textit{ Eur. Phys. J.
  C} \textbf{ 71} (2011) 1554,
  \href{http://dx.doi.org/10.1140/epjc/s10052-011-1554-0}{\doi{10.1140/epjc/s10052-011-1554-0}},
  \href{http://www.arXiv.org/abs/1007.1727}{\texttt{arXiv:1007.1727}}.
  [Erratum: \DOI{10.1140/epjc/s10052-013-2501-z}].

\bibitem{Drewes_2022}
\hrefCMSnoop {}{M.~Drewes, J.~Klari{\'c}, and J.~L{\'o}pez-Pav{\'o}n, ``New
  benchmark models for heavy neutral lepton searches'',} \textit{ Eur. Phys. J.
  C} \textbf{ 82} (2022) 1176,
  \href{http://dx.doi.org/10.1140/epjc/s10052-022-11100-7}{\doi{10.1140/epjc/s10052-022-11100-7}},
  \href{http://www.arXiv.org/abs/2207.02742}{\texttt{arXiv:2207.02742}}.

\end{thebibliography}\endgroup
\cleardoublepage \appendix\section{The CMS Collaboration \label{app:collab}}\begin{sloppypar}\hyphenpenalty=5000\widowpenalty=500\clubpenalty=5000
\cmsinstitute{Yerevan Physics Institute, Yerevan, Armenia}
{\tolerance=6000
A.~Hayrapetyan, A.~Tumasyan\cmsAuthorMark{1}\cmsorcid{0009-0000-0684-6742}
\par}
\cmsinstitute{Institut f\"{u}r Hochenergiephysik, Vienna, Austria}
{\tolerance=6000
W.~Adam\cmsorcid{0000-0001-9099-4341}, J.W.~Andrejkovic, T.~Bergauer\cmsorcid{0000-0002-5786-0293}, S.~Chatterjee\cmsorcid{0000-0003-2660-0349}, K.~Damanakis\cmsorcid{0000-0001-5389-2872}, M.~Dragicevic\cmsorcid{0000-0003-1967-6783}, P.S.~Hussain\cmsorcid{0000-0002-4825-5278}, M.~Jeitler\cmsAuthorMark{2}\cmsorcid{0000-0002-5141-9560}, N.~Krammer\cmsorcid{0000-0002-0548-0985}, A.~Li\cmsorcid{0000-0002-4547-116X}, D.~Liko\cmsorcid{0000-0002-3380-473X}, I.~Mikulec\cmsorcid{0000-0003-0385-2746}, J.~Schieck\cmsAuthorMark{2}\cmsorcid{0000-0002-1058-8093}, R.~Sch\"{o}fbeck\cmsorcid{0000-0002-2332-8784}, D.~Schwarz\cmsorcid{0000-0002-3821-7331}, M.~Sonawane\cmsorcid{0000-0003-0510-7010}, S.~Templ\cmsorcid{0000-0003-3137-5692}, W.~Waltenberger\cmsorcid{0000-0002-6215-7228}, C.-E.~Wulz\cmsAuthorMark{2}\cmsorcid{0000-0001-9226-5812}
% ...
\par}
\vskip\cmsinstskip
\dag:~Deceased\\
$^{1}$Also at Yerevan State University, Yerevan, Armenia\\
$^{2}$Also at TU Wien, Vienna, Austria\\
% ...
\end{sloppypar}
\end{document}